\documentclass[aps,twocolumn,superscriptaddress]{revtex4}
\usepackage[english]{babel}
\usepackage{amssymb,amsfonts,amsmath}
\usepackage[pdftex]{graphicx}
\usepackage[font={footnotesize}]{caption}
\usepackage[official]{eurosym}
\usepackage{mathtools}
\usepackage{natbib}
\usepackage{color}
\usepackage{pdflscape}
\usepackage{afterpage}
\usepackage{subcaption}
\usepackage{rotating}
\usepackage{dcolumn}
\usepackage{bm}
\usepackage{multirow}
\usepackage{color}

\begin{document}

\title{Mobile phone records to feed activity-based travel demand models: MATSim for studying a cordon toll policy in Barcelona }

\author{Aleix Bassolas}
\affiliation{Instituto de F\'isica Interdisciplinar y Sistemas Complejos IFISC (CSIC-UIB), Campus UIB, 07122 Palma de Mallorca, Spain}
\author{Jos\'e J. Ramasco}\affiliation{Instituto de F\'isica Interdisciplinar y Sistemas Complejos IFISC (CSIC-UIB), Campus UIB, 07122 Palma de Mallorca, Spain}
\author{Ricardo Herranz}\affiliation{Nommon Solutions and Technologies, calle Ca\~nas 8, 28043 Madrid, Spain}
\author{Oliva G. Cant\'u-Ros}\affiliation{Nommon Solutions and Technologies, calle Ca\~nas 8, 28043 Madrid, Spain}

\begin{abstract}
Activity-based models appeared as an answer to the limitations of the traditional trip-based and tour-based four-stage models. The fundamental assumption of activity-based models is that travel demand is originated from people performing their daily activities. This is why they include a consistent representation of time, of the persons and households, time-dependent routing, and microsimulation of travel demand and traffic. In spite of their potential to simulate traffic demand management policies, their practical application is still limited. One of the main reasons is that these models require a huge amount of very detailed input data hard to get with surveys. However, the pervasive use of mobile devices has brought a valuable new source of data. The work presented here has a twofold objective: first, to demonstrate the capability of mobile phone records to feed activity-based transport models, and, second, to assert the advantages of using activity-based models to estimate the effects of traffic demand management policies. Activity diaries for the metropolitan area of Barcelona are reconstructed from mobile phone records. This information is then employed as input for building a transport MATSim model of the city. The model calibration and validation process proves the quality of the activity diaries obtained. The possible impacts of a cordon toll policy applied to two different areas of the city and at different times of the day is then studied. Our results show the way in which the modal share is modified in each of the considered scenario. The possibility of evaluating the effects of the policy at both aggregated and traveller level, together with the ability of the model to capture policy impacts beyond the cordon toll area confirm the advantages of activity-based models for the evaluation of traffic demand management policies.
\end{abstract}

\maketitle

\section{Introduction}

Activity-based models (ABMs) offer advantages over more aggregated travel demand models for evaluating policies designed for traffic management. Rather than considering individual trips, ABMs consider the individuals performing them. The trips are seen as the consequence of individuals' desires of performing certain activities. The change of locations 
needed to pass from one activity to the next constitutes thus a trip. The conception of trips linked to individuals make these models more flexible to adapt to different levels of policy application. In addition, it also makes them more appropriate to study the effects of policies aimed at modifying individuals' decisions, e.g., policies of fixed rates for multistage/multimodal trips or different pricing and toll schemes. In spite of the huge potential of these models, their use is mainly restricted to research and there are only few examples of their application for public policy planning. One of the reasons hindering their adoption in real environments is the high level of data requirements. 
Activity-based models require a full diary of activities for each user representing the population of the area of study, which is not always available for the population. Travel surveys, commonly used to obtain such data, provide rich information on travel behaviours but they suffer from major shortcomings. Survey collection is costly and time consuming and depends on users' availability and willingness to answer. This reduces drastically the size of the sample and the frequency with which information can be updated, limiting the information to punctual observations rather than to continuous monitoring. Additionally, the planning needed for survey collection does not allow to obtain data on unexpected events.

The vast amount of spatio-temporal data generated by the use of personal ICT mobile devices provide insights on people's actions and behaviours, bearing valuable information on when and where different actions took place. In particular, geolocated devices such as intelligent transport cards, mobile phones, Bluetooth, global positioning system (GPS), etc. allow for the collection of mobility information requiring minimal or no interaction with the users (collection of passive data). This avoids many of the intrinsic deficiencies of surveys, such as imprecisions on reported time and space, reduces drastically time and cost associated with data collection and provide larger sample sizes. The potential of these new data sources is huge, but it also comes with a number of challenges. On the one hand, we have much larger samples than those obtained from traditional surveys. On the other hand, the data have not been originally produced for the purpose of collecting activity-travel information, and, therefore, it is often noisy and/or biased. The reconstruction of activity and mobility patterns calls thus for the development of ad hoc data analysis methods addressing the specific characteristics of each dataset.
 
In the last decade, several studies, starting from the pioneer work of Marta Gonz\'alez et al.in \cite{Gonzalez} to more recent ones like those of Bagrow and Lin \cite{Bagrow}, Lenormand et al. \cite{Lenormand, Lenormand2}, Louail et al. \cite{Louail,Louail2} and  Picornell et. \cite{Picornell} among others, have been carried out to investigate how data obtained from mobile phone records can be used to characterise people mobility habits and the factors influencing them (see Blondel et al. \cite{Blondel} and Barbosa et al. \cite{Barbosa} for recent reviews). Mobile phone data has also been used for the characterisation of mobility patterns in specific situations such as in the work by Ahas et. al \cite{Ahas} monitoring the movements of work trips in Tallinn,  Calabrese et al. \cite{Calabrese} studying how people move during large social events, Becker et al. \cite{Becker} identifying the residences of workers and the nightlife of Morristown, Song et al. \cite{Song} and Isaacman et al. \cite{Isaacman} identifying locations where people spend most of their time and characterising how they return to them. In the transport sector, research on the use of mobile phone records has been mainly focused on the estimation of aggregated variables related to travel demand, such as travel time \cite{Bar-Gera}, mode \cite{Wang,Doyle} and route \cite{Tettamanti} choice, estimation of origin-destination matrices \cite{Alexander,Caceres2011}, and traffic flows \cite{Caceres2011}.  Trip purpose characterization from mobile phone records has also been approached by different authors \cite{Alexander,Rose, Caceres2008,Steenbruggen,Phithakkitnukoon,Gong}. Most of the work previously discussed belongs to academic research and has not been applied to real-world planning projects \cite{Lee}. 
In terms of integration with simulation models, so far there are only few examples of traffic models fed with demand information (origin-destination matrices) generated from of mobile phone data \cite{Lee}. This is partly because the obtained information does not always meet the requirements of format, level of resolution and completeness.

The main contribution of the work presented here is to show the potential of mobile phone data to generate activity travel dairies to feed an activity-based travel model aimed at evaluating the impact of a traffic management policy. Activity diaries of residents of the metropolitan area of Barcelona are reconstructed from mobile phone records. Such diaries are used as input for the activity-based traffic model MATSim to evaluate the impact of a cordon toll applied to two different areas of the city. Results of the policy are obtained and discussed at an aggregated and at a resident-centric level. The paper is organized as follows: Section \ref{sec:description and implementation} describes the model and its implementation for the region of the study, including model calibration and validation. Section \ref{sec:case study} presents details of the case study, the cordon toll implementation. In Section \ref {sec:results and discussion} the results of the calibration and validation process as well as of the policy implementation are discussed. Finally, Section \ref{sec:conclusions} presents the main conclusions of the work, divided into conclusions of the model implementation process and conclusions derived from the policy implementation.

\section{ Model description and implementation}\label{sec:description and implementation}

\begin{figure}
	\begin{center}
		\includegraphics[width=8cm]{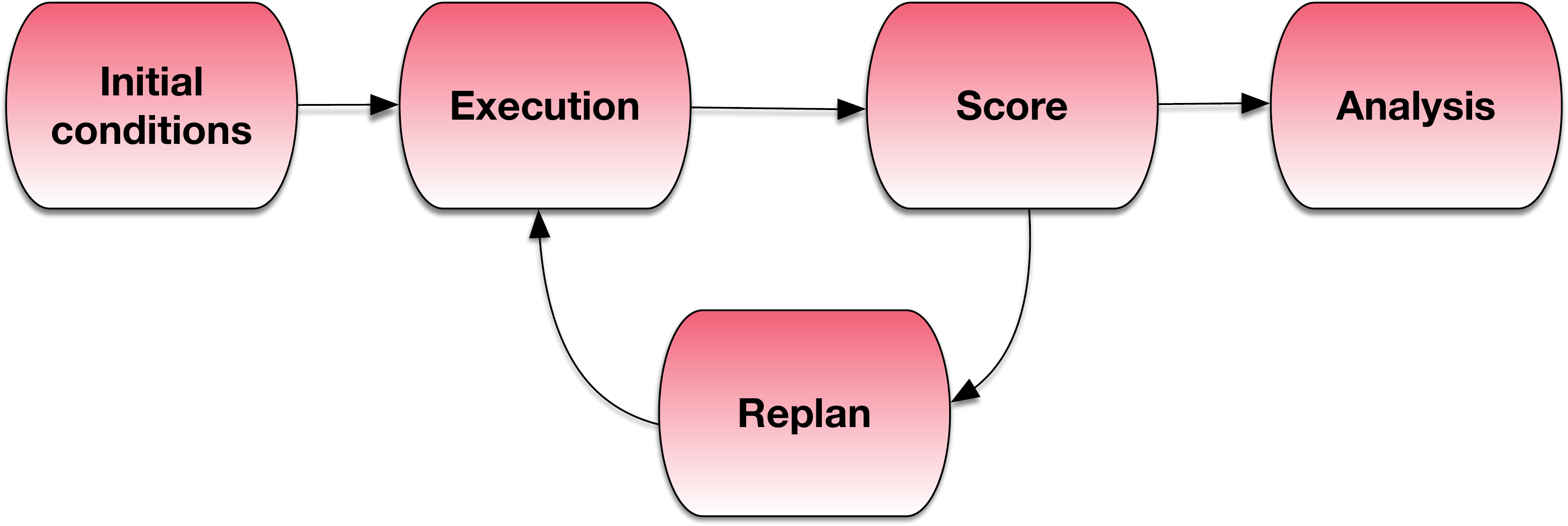}
		\caption{Sketch of MATSim main flow diagram (see http://www.matsim.org/about-matsim for detailed information on the structure, the implementation and how to run MATSim). \label{matsim}}
	\end{center}
\end{figure}

\subsection{MATSim: Multi-Agent Transport Simulation}\label{MATSim_model}
 
MATSim is an activity-based multi-agent simulation framework used to simulate traffic flows and the possible congestion associated to them. In its typical configuration the simulations cover one generic day. In MATSim, travel demand and travel flows are generated by agents performing their daily activities. Each agent attempts to maximise the utility of its daily activity schedule while competing for space-time slots with all other agents on the transportation infrastructure. Agents maximise their utility by minimising the total travel cost, $V_{trav} = \sum_{j} V_{trav,j}$, and maximising the time spent performing activities, $\sum_{i} V_{act,i}$. The index for activities $i$ runs between $1$ and $N$, the total number of activities on schedule for the day. The trips occur between activities and, therefore, $j$ goes from $1$ to $N-1$. $V_{act,i}$ is a logarithmic utility function associated with the time spent performing activity $i$ and it is defined as
\begin{equation}
V_{act,i}=\beta_{dur}+t_{typ}+ln\left( \frac{t_{dur,i}}{t_{0,i}} \right),
\end{equation}
where $\beta_{dur}$ is the marginal utility of performing an activity, $t_{typ}$ the activity typical duration, $t_{dur,i}$ the actual duration of the activity $i$, and $t_{0,i}$ the minimal duration after which the utility starts to be positive. Although the possibility could be easily implemented (see \cite{MATSim_book}), in this work early and late arrivals to activities are not penalised. The trip utility function, $V_{trav,j}$, is associated to the cost (time and money) of travelling from one activity $j$ to the next $j+1$ and it has the following linear expression: 
\begin{align}\label{eq4}
V_{trav,j}= C_{mode(j)}+ \beta_{trav,mode(j)} \,t_{trav,j}+\beta_m \, \Delta m_{j} \nonumber \\+ \left( \beta_{d,mode(j)} 
+ \beta_{m} \, \gamma_{d,mode(j)}\right) \,d_{trav,j}, 
\end{align} 
where $C_{mode}$ stands for a mode-specific constant, $\beta_{trav,mode}$ for the direct marginal utility of time spent travelling by mode, $t_{trav}$ for the time taken to travel from activity $j$ to activity $j+1$, $\beta_{m}$ is the marginal utility of money, $\Delta  m$ the change in monetary budget caused by fares or tolls, $\beta_{d,mode}$ is the marginal utility of distance (normally negative or zero), $\gamma d$ is the mode specific monetary distance rate (negative or zero), and $d_{trav}$ the distance travelled between activity $j$ and activity $j+1$.

Users minimise the travel cost by taking decisions on the route choices, transportation modes and activities' sequence and timing, based on previous experiences (replaning process). Although some implementations of MATSim allow the modification of the location of activities in the replaning process (\cite {Horni:2009}), this is not the case of the work presented here. Activity diaries are modified iteratively until the system finds a stationary state, i.e., further changes in the activity diary do not produce qualitative changes in the total score of the system (sum of the individual utilities of each of the participating agents). The iterative process is depicted in Figure \ref{matsim}. More details about how to implement and run MATSim can be found at \cite{MATSim_book}.
 
The minimal required inputs to run the model can be summarised as: i) travel demand, consisting of a population of agents and their plan of activities. This should contain a list of all activities to be performed by each agent during the day and the time and location where they take place. As initial condition, it is also required a transport mode for the trips that along with other details of the plan will be later subjected to changes during the simulation to improve agents utility. Due to the high cost in terms of computational time and memory, typically only a representative sample of the total population is simulated. And, ii) supply network, consisting of the full information of the road and the public transport network of the studied area.   

\subsection{Model implementation}

\subsubsection{ Region of study }

Barcelona is the second largest and the densest city of Spain. The metropolitan area is conformed by $36$ municipalities and according to the Spanish national statistical office, INE, in $2017$ it had a population of $3,247,281$ inhabitants. The municipality of Barcelona is the core of the area and it concentrates around one half of the population with $1,620,809$ inhabitants. Jobs, retail and cultural offer make the municipality of Barcelona a trip attractor. Actually, up to $27 \%$ of the trips registered in the Barcelona municipality are due to residents of the rest of the metropolitan area. The metropolitan area is an official entity that manages a public transportation network composed of four transport  modalities: bus, metro, train and tramway \cite{AMB}. The network is organized in $6$ concentric fare zones centred at Barcelona municipality, and covering virtually the whole province. The majority of traffic is concentrated in the first fare zone, whose residents are responsible for $85 \%$ of the trips in the Barcelona municipality. The road network, on the other hand, has three important landmarks: the central system, consisting of Barcelona Rondas, which is the ring around the city, and their distribution accesses to outer roads, the main axes of distribution and territorial structure, and the bypass defined by B-30 road. Barcelona Rondas are one of the most important infrastructures of the metropolitan road system. These are high capacity rings with annual average daily traffic intensities exceeding $166,000$ vehicles per day, surrounding the
city without interfering with the urban interior network. The Rondas are divided into two turn-offs: Ronda Litoral, seaside, and Ronda de Dalt, mountain side. They have a total length of $36$ $km$, of which $8.5$ are tunnels (See Figure \ref{transportnet}). 

\subsubsection{Assumptions and constraints in the model setting}

Taking into account the characteristics of the region as well as the available data and the computational requirements of the model, the following constrictions are applied to the model settings:
\begin{itemize}
\item The model is implemented for the first fare zone of Barcelona;
\item Only trips performed by residents in the area considered are simulated: passing by trips and freight trips are excluded from the study;
\item Four different transport modes are considered: private car, public transport (including  train, underground, tramway and bus), bike, and walking;
\item  Only travellers older than $15$ years old are included in the simulations. This obeys to two reasons: i) the use of mobile phones is not fully spread among the population below that age. And ii), the official statistics used to calibrate the model only consider the mobility of people older than $15$ years. Even though children are not explicitly included, some of their trips are indirectly contemplated (especially for the youngest ones) since, for instance, bringing them to school or to after-school activities is already part of the parents' activity diary. 
\item Due to the large number, $\approx 2,300,000$, of inhabitants older than $15$ years in the studied area and the high computational requirements in terms of memory and simulation time, a subpopulation of agents formed by  $10 \%$ of the total population is taken as a representative sample.
\item Three types of activities are considered: home, work and other.
\item The base year for the study is 2014, since it corresponds to the time of the phone call records available.
\item The location of each activity is fixed: even when some MATSim implementations allow activity locations to be modified by adding an extra layer with land use information, this option is not implemented in this work.
\end{itemize}

\subsubsection{Travel demand: Population generation}\label{travel_demand}

The use of transport infrastructures cannot be well framed if the demand considered is not realistic, hence the agents should be representative of the full population. Different socio-demographic characteristics such as age, gender and residence place influence the way people move \cite{Lenormand3}. This implies that the simulated population should be representative of the whole population not only in terms of a variety of mobility/activity patterns, e.g., number and length of trips, activities' locations, etc., but also of the socio-demographic characteristics. 

In this work diaries of activity as well as residence place are reconstructed from Call Detail Records (CDRs), and  census information is used to up- or down-scale the sample to the desired $10 \%$ of the population, reproducing the official aggregated age and gender statistics at census track level. A brief description of the CDRs main characteristics and an explanation of the process followed to build a representative sample from both data sources is given next.\\

\paragraph{Call Detail Records}\mbox{}\\

CDR data is produced every time a mobile phone interacts with the network through a voice call, a text message or an Internet data connection. The records contain information about the time and position of the tower to which the device connects. This provides an indication of the geographical position of the user at certain moments along the day with a time and space granularity depending on the level of activity of each user and on the technology deployment of the service provider. In this work, anonymised Call Detail Records for the period of October-November 2014 provided by Orange Spain are used to reconstruct activity diaries of users residing at the study region. Orange is currently the second largest mobile network operator in Spain, with a market share of around $35\%$ in Catalonia. This ensures a large sample size in the area studied.

The CDRs used contain the following information: a unique anonymised identifier per user, the time when an interaction with the network occurred and the geographic coordinates of the  tower at which the device is connected at that particular moment. There are two limitations that should be taken into account: i) There is a time window between two consecutive events, typically between $20-30$ minutes for the most active users. And ii), the tower position is not the exact position of the user. The space granularity observed in the CDRs in the region of study is of around $100-200$ meters. Besides CDRs, also some extra information such as age and gender is known for the users.\\

\paragraph{ Diaries reconstruction from CDRs}\mbox{}\\ 

For diaries reconstruction, only the most active users are kept leaving a sample corresponding to $15\%$ of the total population. Once a useful sample is selected, space is divided into Voronoi areas according to the base transceiver station (BTS) tower position. Using longitudinal information recorded during the whole October-November period an activity profile at a Voronoi area level is built for each user. The frequency of appearance and the time and length of stay in each area are used to identify the user's main locations (home, work, other)  as well as the typical higher-lower activity times. This constitutes the core of the  activities profile of the user. Once a profile is built,  the diary of a sample day is built for each user. All the visited locations in the sample day are identified, including locations different from the common ones so as to reconstruct non recurrent trips. According to the user's activities' profile, a probability function is used to adjust the start and end time of each activity/trip. The start time of activity i is chosen to take place between the last register corresponding to activity i-1 plus the estimated travel time between locations i and i-1, and the first register at activity i. The end time of activity i-1 is thus the start time of activity i minus the estimated travel time. Next, based on the identified home location non-residents of the Metropolitan area of Barcelona are filtered out. Finally, the activity-travel diaries are extrapolated to the desired percentage ($10\%$) of the total population.\\

\paragraph{ Sample expansion}\mbox{}\\ 

The residence records ({\it padr\'on}) of 2014 are used to obtain a sample representative of the $10\%$ of the total population. The expansion is performed at the geographical level of census tract. For each tract, the information of the {\it padr\'on} is aggregated in six categories separating the population by gender and in the following age ranges: $0-15$ , $16-64$ and over $64$ years. Since the CDR data is restricted to individuals over $15$, only the last two age groups are considered for the expansion.

Residents located at a given Voronoi area are assigned to one of the census tracts intersecting it or one of its neighbouring areas. The assignation is made with a probability directly proportional to the square of the population of the census tract and inversely proportional to the square of the number of users already assigned to that tract. The assignation process ensures a local homogeneous sample density among neighbouring census tracts. 

For each age-gender category in every census tract, the sample is either expanded (agents are "cloned") if the sample under-represents the preestablished threshold of $10\%$ of the population of the given category, or reduced (agents are randomly selected) if its number exceeds the threshold per category. In this way, the distribution of users by category matches that provided by the population records in every tract. For the expansion, given the gender $g$, the age group $a$ and the tract $i$, each agent is copied $n$ times where $n$ is given by the integer part of the ratio between the $10\%$ of the population, $P_i^{a,g}$, and the sample size for the given category and census tract $S_i^{a,g}$ minus one, $n = [P_i^{a,g}/S_i^{a,g}]-1$. Finally, some agents, $P_i^{a,g} - [P_i^{a,g}/S_i^{a,g}]\, S_i^{a,g} $, are randomly selected to be cloned one extra time.

\begin{figure}
	\begin{center}
		\includegraphics[width=8cm]{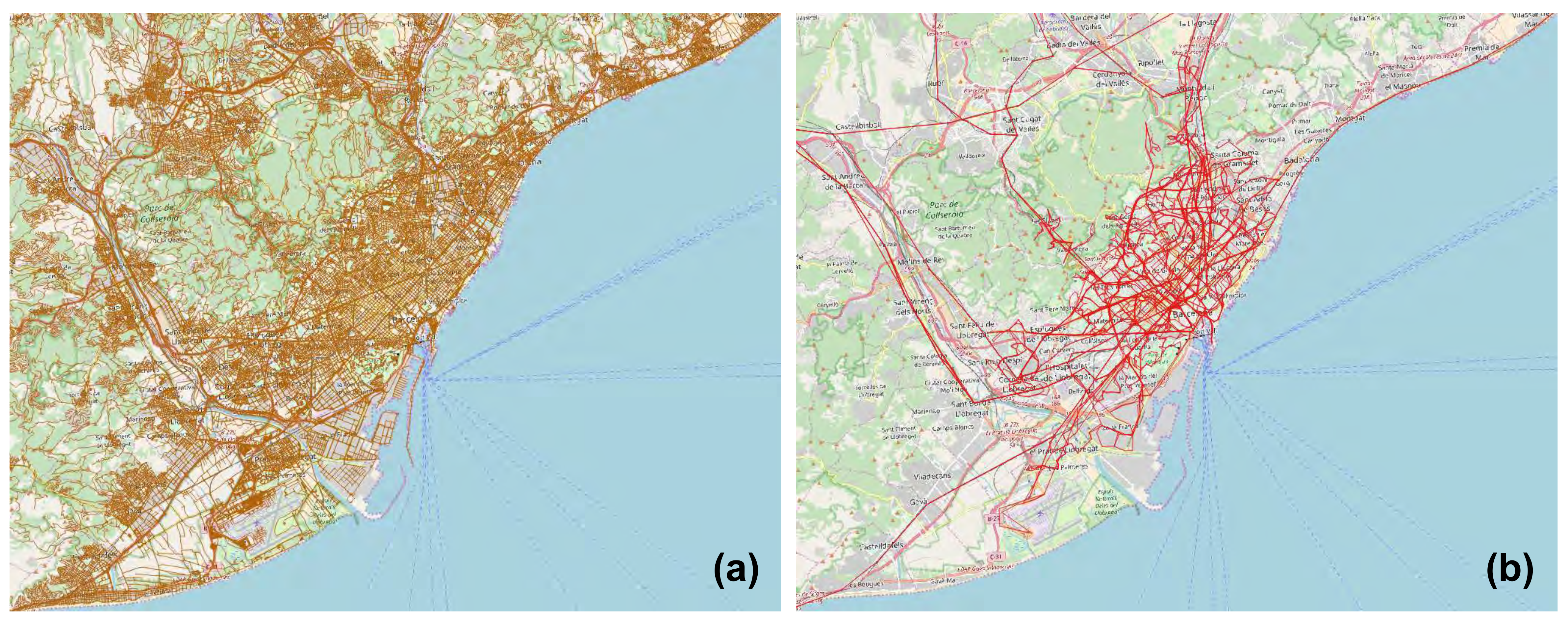}
		\caption{(a) Road network and (b) public transport network as obtained from Open Transport Maps (http://opentransportmap.info/) and Open Street Map (www.openstreetmap.org). The layout of the maps was obtained from Open Street Map.  \label{transportnet}}
	\end{center}
\end{figure}

\subsubsection{Supply network: Metropolitan area of Barcelona transport infrastructure}

\paragraph{Road network }\mbox{}\\

The data used to build the road network was obtained from Open Transport Maps (\url{http://opentransportmap.info/}) and Open Street Map (www.openstreetmap.org). The information provided  for each link conforming the road network includes: coordinates for start and end nodes, traffic directions allowed, type of the road (from highways to pedestrian streets), free speed in km/h, road capacity per link, and number of lanes per link. This information was treated in order to build the network for MATSim implementation: link length was calculated using the link's start and end coordinates, links capacity was scaled down to the $10\%$ of actual capacity to take into account the sampling of the population.
Additional links were generated to account for bi-directional links, since MATSim only accepts unidirectional links. The new links inherit all properties of the original ones but their start and end nodes are swapped. Pedestrian streets and bike lanes as well as unconnected and loop links (those where start and end nodes are the same) were removed. For those links with no information of free speed, capacity and/or number of lanes, an average value was imputed according to the road type. The resulting network, depicted in Figure \ref{transportnet}a, contains $17,690$ road links and $9,217$ nodes.\\

\paragraph {Public transport network}\mbox{}

\subparagraph{Bus and metro}
Information about stops, routes, schedules and departures has been obtained from the public information available at the Barcelona Open Data platform (\url{http://opendata.bcn.cat/opendata/en}). The bus network of Barcelona has $192$ lines and $2,464$ stops. The metro network is composed of $22$ lines and $139$ stops. 
\subparagraph{Train and tram}
Train and tram networks data was obtained by querying  web sites of the public companies managing the service (Renfe \cite{renfe} and Rodalies de Catalunya \cite{rodalies}) and extracting structured data from them.
The Barcelona metropolitan rail network is composed of $38$ lines and $181$ stops. The tram has two networks with a combined total of $12$ lines and $56$ stops. 

Although specific code was written to download and parse the data, the process was the same for all the sources: to detect the lines, to assign a unique identifier to each line and to extract the stops in both directions (and give them a unique identifier). It has to be noted that in some cases the stops are not the same in both directions. For every pair of consecutive stops along every line, we extract the duration of the journey and for every head stop (in both directions) we also get the time the convoy starts the journeys. The resulting public transport network is depicted in Figure \ref{transportnet}b.

\subsection{Model Calibration and validation} 

\subsubsection{Model calibration}

The calibration consists of adjusting the values of the parameters described in section \ref{MATSim_model} such that certain outputs of the simulation (in our case, modal split) are consistent with baseline values. 
The output of MATSim contains the resulting route and mode used by each agent during his/her displacements between activities. This allows us to extract the modal split and to compare it with the one in the EMEF survey ({\it Encuesta de Mobilitat en Dia Feiner} 2014 \cite{emef}). This is done at two different geographical scales: the full metropolitan area and the shire of Barcelona, called Barcelon\`es, which includes the municipality of Barcelona and four other municipalities. The parameters of MATSim were set in such a way that simulated modal split matched the observed ones in the two areas at the same time. The results from the calibration are shown in Section \ref{sec:validation_results}.

\subsubsection{Model validation} 

Traffic counts of some of the main entrances of the city were used to validate the results of the model once calibrated. This data was obtained from the web portal of statistics of the Barcelona City Council \cite{AjuntamentdeBarcelona}. 
Only those links with a traffic volume higher than $10,000$ vehicles per day were considered. Traffic statistics at the different links reported by MATSim are obtained from the agents final travel diary, hence the "traffic" volumes in each road corresponds to people's car trips rather than real number of cars. Since a car may be occupied by more than one person, people's car trips should be converted into actual number of cars. The conversion is made by applying an occupation factor. The value of this factor may vary depending on the road, the time of the day and the trip purpose. In this work, we consider the occupation factor reported by the city's council in the sustainable urban mobility plan (SUMP) \cite{SUMP}, which is $1.25$. For the validation exercise, we have considered possible values ranging from a minimum simulated car volume, corresponding to an occupation factor equal to this $1.25$, and a maximum volume corresponding to a unit occupation factor. The validation process consists in corroborating that for each road for which information was available, the observed number of cars passing by the road falls within the interval of possible values obtained by MATSim for the given road. Details on the observed and simulated number of cars passing by the main roads are given in Section \ref{sec:validation_results}.

\begin{figure}
	\centering 
	\includegraphics[width=8cm]{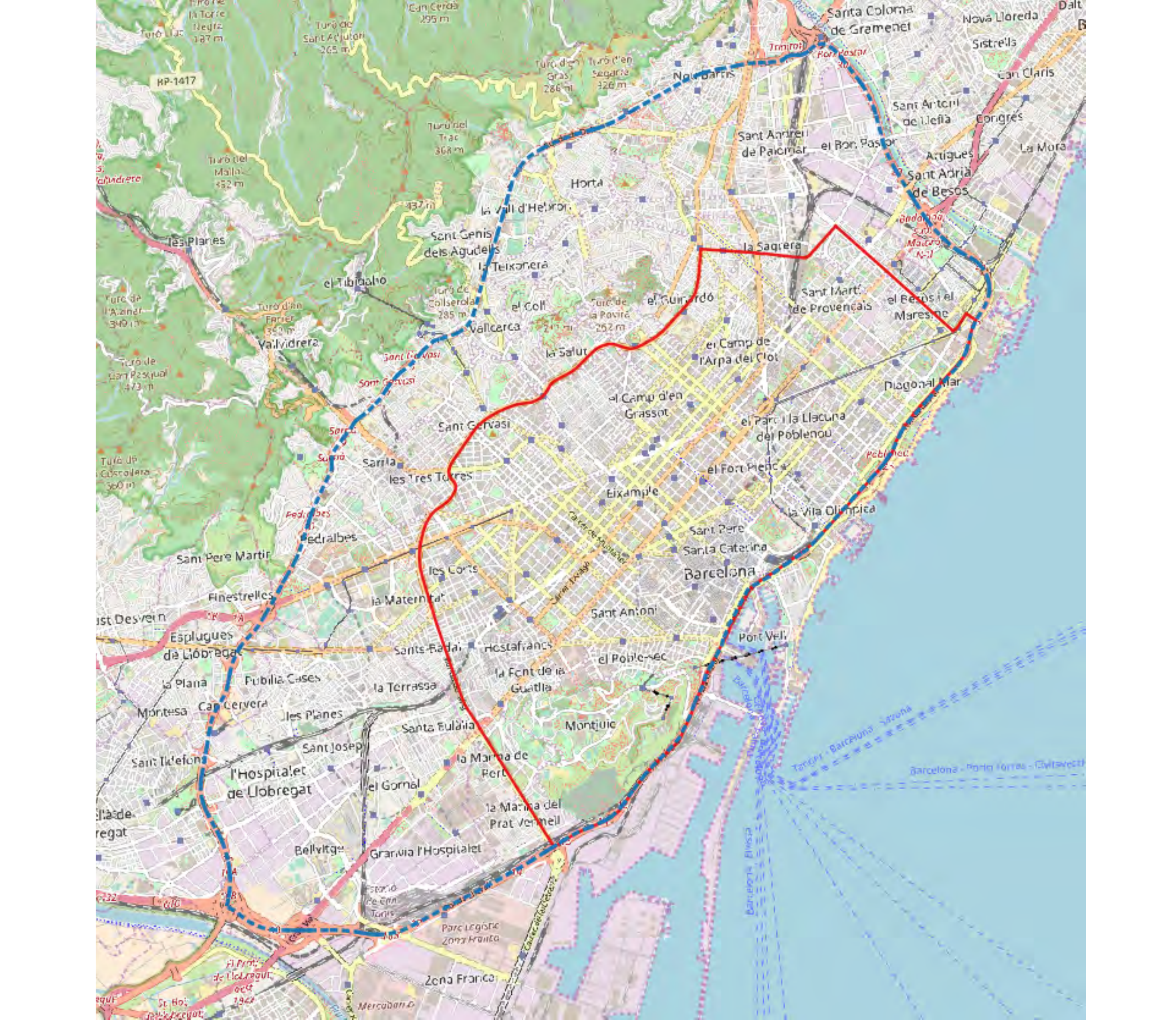}
	\caption{ Inner toll area delimited by the red solid line and Peripheral toll area delimitated by the blue doted line. The maps and the layout were produced with Open Street Map.  }\label{fig:toll_areas}
\end{figure}

\section{Case of study: Cordon toll policy}\label{sec:case study} 

A case study has been designed to accomplish the twofold objective of this work: i) test the potential of using ICT data sources, in particular mobile phone records, to reproduce mobility patterns able to feed, the information demanding ABM, and ii) demonstrate the advantages of Activity Based Models to assess the impact of different policies for traffic management. The application of a cordon toll has been simulated in the central area of the city, a policy that has been applied in other places to reduce the use of private cars and decrease congestion and pollution, both major issues in Barcelona too.

The toll policy is implemented as follows: a ring delimiting the toll area is set. A fee is charged every time a user enters the toll area by car. Trips performed by public transport and soft-modes (walk and bike) are exempted of charge. Once in the area, the user can circulate freely of charge regardless the time or distance travelled inside. The charge is applied to all users independently of their residence place, i.e. residents of the toll area are not excluded from the payment when they enter into it.
Two different areas are considered: one comprising the city centre (red line in Figure \ref{fig:toll_areas}), and a peripheral area surrounded by the Ronda Litoral and the Ronda de Dalt (blue dotted line in Figure \ref{fig:toll_areas}). In both cases, circulation within the Rondas delimiting the charging zone is free of charge.

\subsection{Policy implementation in the MATSim model} 

The cost of entering the tolled area by car reflects in the user's utility function. A list of links entering from the non-toll to the toll zone is provided to MATSim. Every time a traveller uses a link connecting a point outside the toll area with a point insider the toll area, the cost is charged in the utility function via the $\Delta m_{j}$ described in Equation \eqref{eq4}.

\subsection{Policy scenarios} 

Four different policy options are tested:
\begin{enumerate}
\item Fixed congestion rate along the day;
\item Congestion charge applied only during the morning (08:00-10:00) and afternoon (16:00-20:00) rush hours;
\item Congestion charge applied only during the morning peak;
\item Congestion charge applied only during afternoon peak.
\end{enumerate}
All-day congestion charge is expected to affect all kinds of trips, while congestion charge for the peak hours is expected to affect mainly commuting trips. The asymmetric congestion charges for morning or afternoon peak are expected to affect only one direction commuters: congestion charge in the morning peak will mainly affect those commuters travelling from outside to the centre of Barcelona (toll area), while congestion charge applied during the afternoon peak will mainly affect commuters living in the city centre of Barcelona (inside the toll area) and working outside the toll area, since they will have to pay it in their way back home during the afternoon peak.

The first policy option is applied for both peripheral and inner areas. Charges of  $2$ \euro{}, $5$ \euro{} and $10$ \euro{} are considered. For the policy options 2 to 4, only a $10$ \euro{} charge applied to the inner area is tested.
\begin{figure}
	\begin{center}
		\includegraphics[width=8cm]{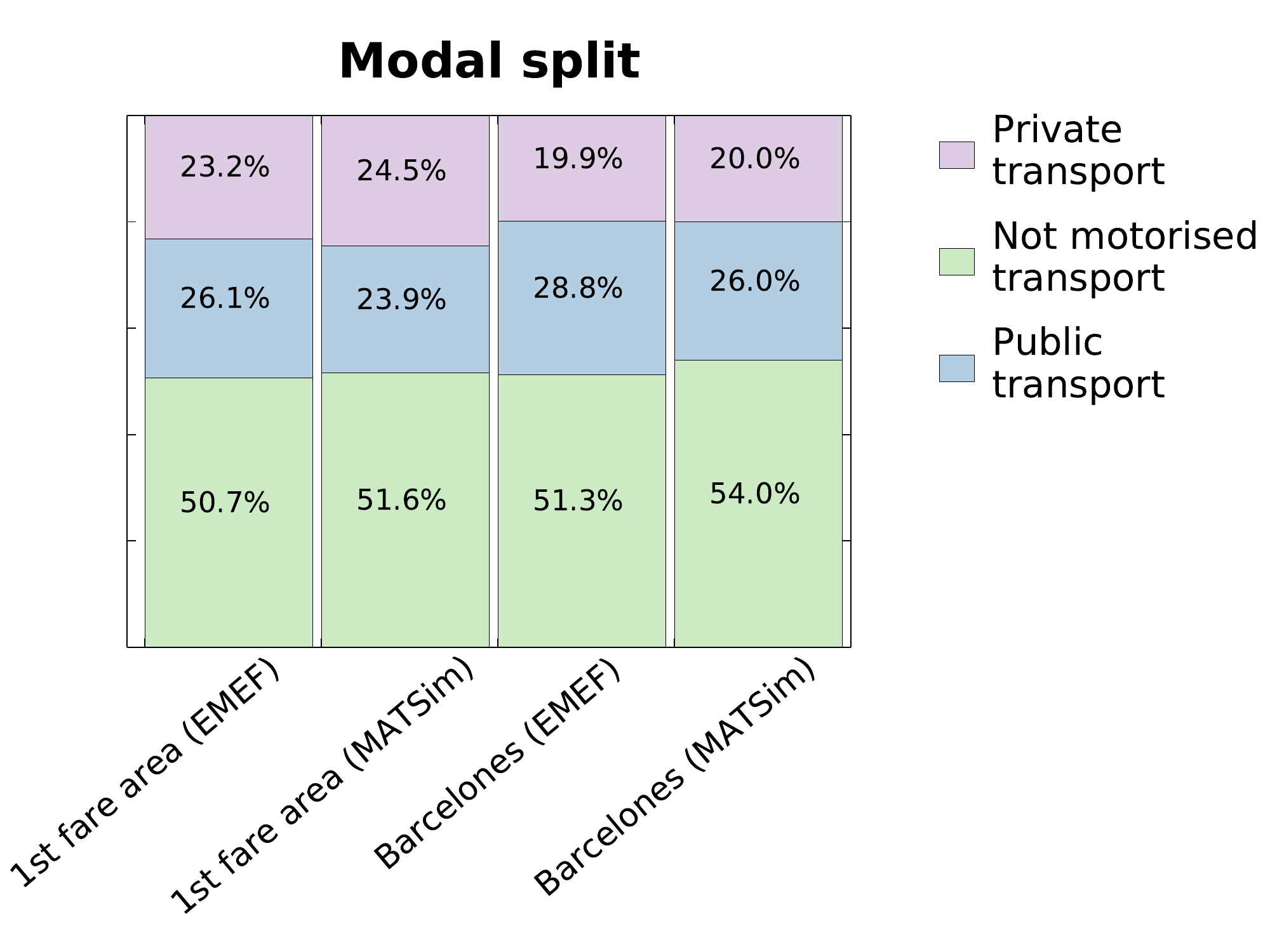}
		\caption{Comparison between observed and simulated modal split for the Metropolitan area and the Barcelon\`es (inner shire of Barcelona).\label{Calibration modalsplit: plot}}
	\end{center}
\end{figure}

\section{Results and discussion}\label{sec:results and discussion}

\subsection{Calibration and validation results}\label{sec:validation_results}

\subsubsection{Calibration} 
A comparison between the modal split reported by the EMEF and the one obtained after the model calibration for both the first fare zone of the Barcelona metropolitan area and for the Barcelon\`es are shown in Figure \ref{Calibration modalsplit: plot}. Observed and simulated car mode shows a difference of less than $1.5\%$. The parameter values leading to this modal split are reported in Table \ref{parameterValues}.

\begin{figure}[b]
	\begin{center}
		\includegraphics[width=8cm]{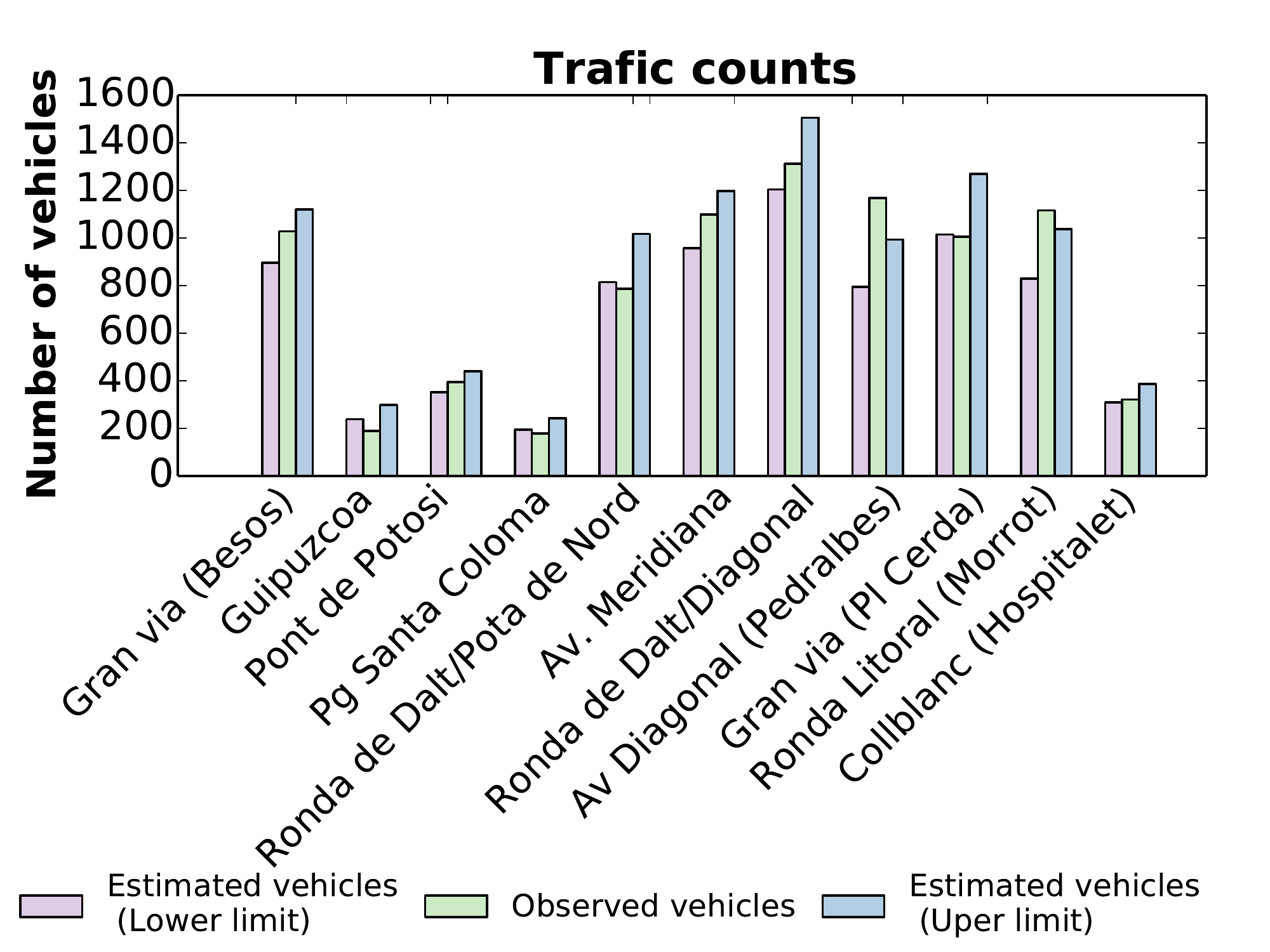}
		\caption{Comparison between vehicle counts observed and obtained from simulation. \label{traffic counts: plot}}
	\end{center}
\end{figure}

\begin{table}
\scriptsize
\begin{center}
\begin{tabular}{|c|c|}
\hline
Parameter & Value \\ \hline
$\beta_{dur}$ &  $6$ \\ \hline
$C_{mode(car)}$ & $-13$ \\ \hline
$C_{mode(walk)}$ &  $0$ \\ \hline
$C_{mode(bike)}$ & $-22$ \\ \hline
$C_{mode(public transport)}$ & $-4$\\ \hline
$\beta_{trav,mode}$ & $-6$ for all modes \\ \hline
$\beta_{d,mode}$ &  $0$ for all modes \\ \hline
$\gamma_{d,car}$ & $-7.7 \times 10^{-5}$\\ \hline
$\gamma_{d,mode}$ &  $0$  for all modes but car \\ \hline
$\beta_{m}$ & $ 1$ \\ \hline
\end{tabular}
\end{center}
\caption{Parameter values}
\label{parameterValues}
\end{table}

\subsubsection{Validation} 
Observed and simulated number of cars passing by the main roads are shown in Table \ref{table3} and Figure \ref{traffic counts: plot}. Note that traffic counts were not used during the calibration process, i.e., the resulting travel counts from the simulation have not been influenced by the observed ones in any way. We can see that observed vehicle counts of most of the main roads lay between the minimum and maximum counts estimated from the simulation. For an occupation factor of $1.25$, $9$ out of the $11$ roads compared in Table \ref{table3} present a GEH statistic below $10$. The Ronda Litoral and Avenida Diagonal are the two outliers having more observed vehicles than the maximum estimated. A reason for this is that we are not considering freight and passing by traffic. Ronda Litoral is one of the main access for freight traffic to the premisses of the port of Barcelona and to Mercabarna, the main logistic centre of the city. Avenida Diagonal is the main road crossing the city from West to East and hence it has a lot of passing by traffic. 

\begin{table}[b]
\scriptsize
\begin{center}
\begin{tabular}{ | l | l | l |  }
\hline
Road & Observed counts & Simulated counts\\
     &		       & lower-upper limit\\ \hline
Gran via & $102,788$ & $89,616-112,020$\\ 
(Besos)  &		       & \\ \hline
Guipuscoa& $18,940$ & $23,864-29,830$ \\ \hline
Pont de Potosi  & $39,435$ & $35,168-43,960$\\ \hline
Pg Santa Coloma & $17,835$ & $19,456-24,320$\\ \hline
Ronda de Dalt/& $78,683$ & $81,416-101,770$\\
Pota de Nord&		       & \\ \hline
Av. Meridiana& $109,885$ & $95,784-119,730$ \\ \hline
Ronda de Dalt/& $131,159$ & $120,464-150,580$\\ 
Diagonal &		       & \\ \hline
Av Diagonal & $116,766$ & $79,472-99,340$\\ 
(Pedralbes) &		       & \\ \hline
Gran via &$100,546$ & $101,510-126,888$\\ 
(Pl Cerda) &		       & \\ \hline
Ronda Litoral & $111,548$ & $82,952-103,690$\\ 
(Morrot) &		       & \\ \hline
Collblanc &$32,155$ & $30,920-38,650$\\ 
(Hospitalet)&		       & \\ \hline		
\end{tabular}	
\caption{Road counts comparison between simulated results and official counts}
\label{table3}
\end{center}
\end{table}

\subsection{Policy results} 

Results are analysed at two different levels: at an aggregated level, e.g., total or average travel time, modal split, total number of car trips, etc., and from a user centric perspective, e.g., the residents most affected by the different levels of the policy application.

\subsubsection{All-day toll charge aggregated results}\label{all_day}
\begin{figure}
	\begin{center}
		\includegraphics[width=7cm]{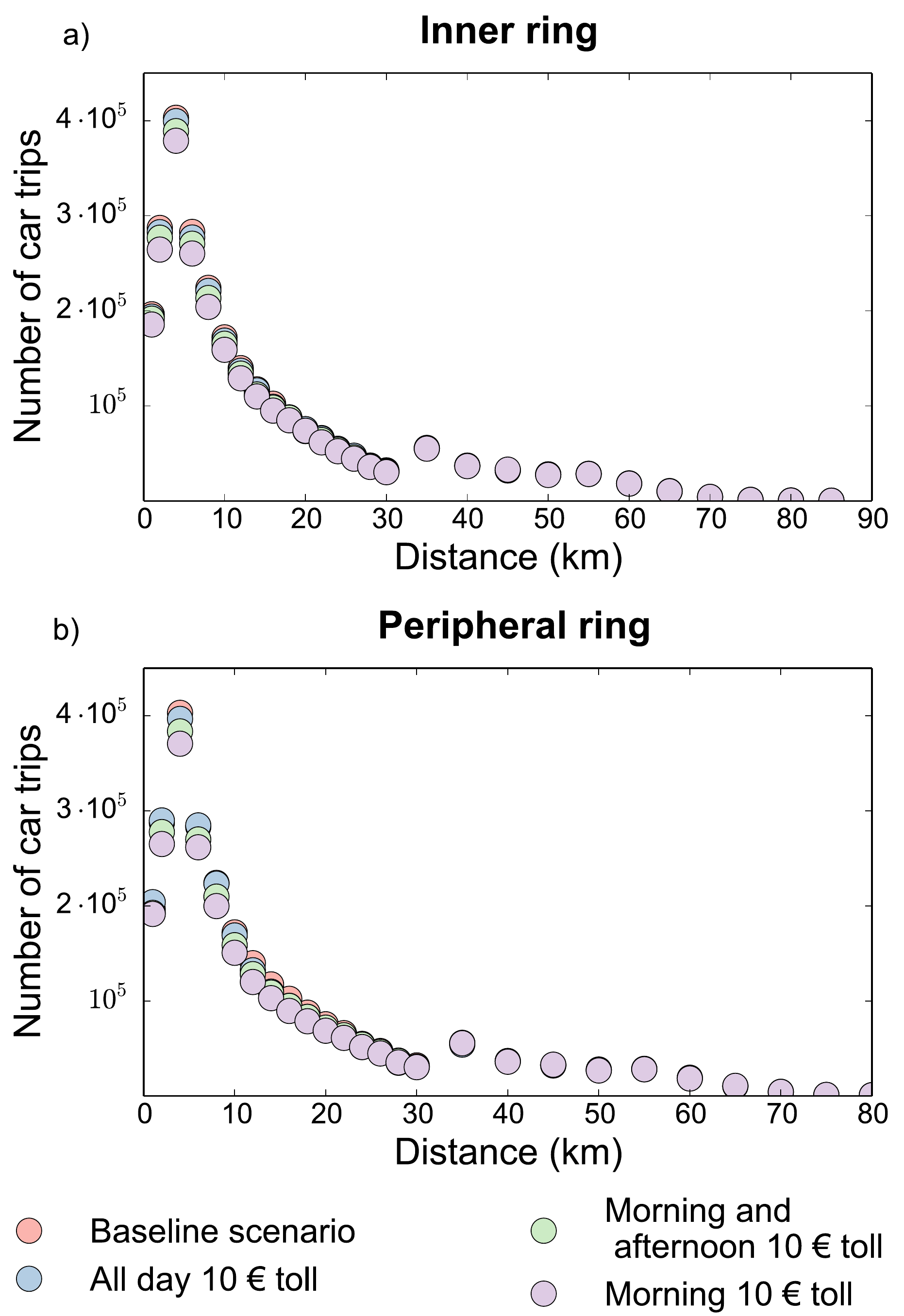}
		\caption{ Number of car trips per travelled distance. a) Inner toll zone and b) peripheral toll zone  \label{ fig:car_trips_distance} }
	\end{center}
\end{figure}

The most relevant results from the simulation of the different scenarios are discussed here. Details and numerical values of the aggregated results for all the different scenarios can be found at Tables \ref{table_peri} and \ref{table_inner} in Appendix A.

For both inner and peripheral rings, results show that the number of car trips decreases with a growing toll price. Car trips reduction also intensifies with the area covered by the ring toll: the bigger the area the less car trips are observed (See Figure  \ref{ fig:car_trips_distance}). A reduction of  $1.2\%$, $3.2\%$, $6\%$ of car trips for the inner ring and of $1.7\%$, $4.4\%$, $8\%$ for the peripheral ring are observed for $2$ \euro{}, $5$ \euro{}  and $10$ \euro{}  toll charge, respectively. Car use decrease occurs mainly for short distance trips (Figure \ref{ fig:car_trips_distance}). This may be due to a low connectivity of the public transport at long distances. If the public transport is not optimum, driving is still a good option even with a tolling scheme. We discuss next only the results of the $10$ \euro{}  charge (for the results of the other charges see Tables \ref{table_inner} and \ref{table_peri} in Appendix A). Car use decrease not only for trips entering the toll zone but also for trips exiting the zone or trips starting and ending inside or outside the toll zone. As expected, the higher reduction occurred for trips entering and exiting the toll zone with $18\%$  less cars crossing the ring toll in any direction for both inner and peripheral rings cases. A reduction of $9.8\%$ for the inner ring and of $7\%$ for the peripheral ring resulted for car trips starting and ending inside the toll zone. For car trips starting and ending outside the toll zone, a reduction of $1.8\%$ and of $2.5\%$ is observed for the inner and peripheral ring tolls respectively (Tables \ref{table_inner} and \ref{table_peri}).

\begin{figure}
	\begin{center}
		\includegraphics[width=8cm]{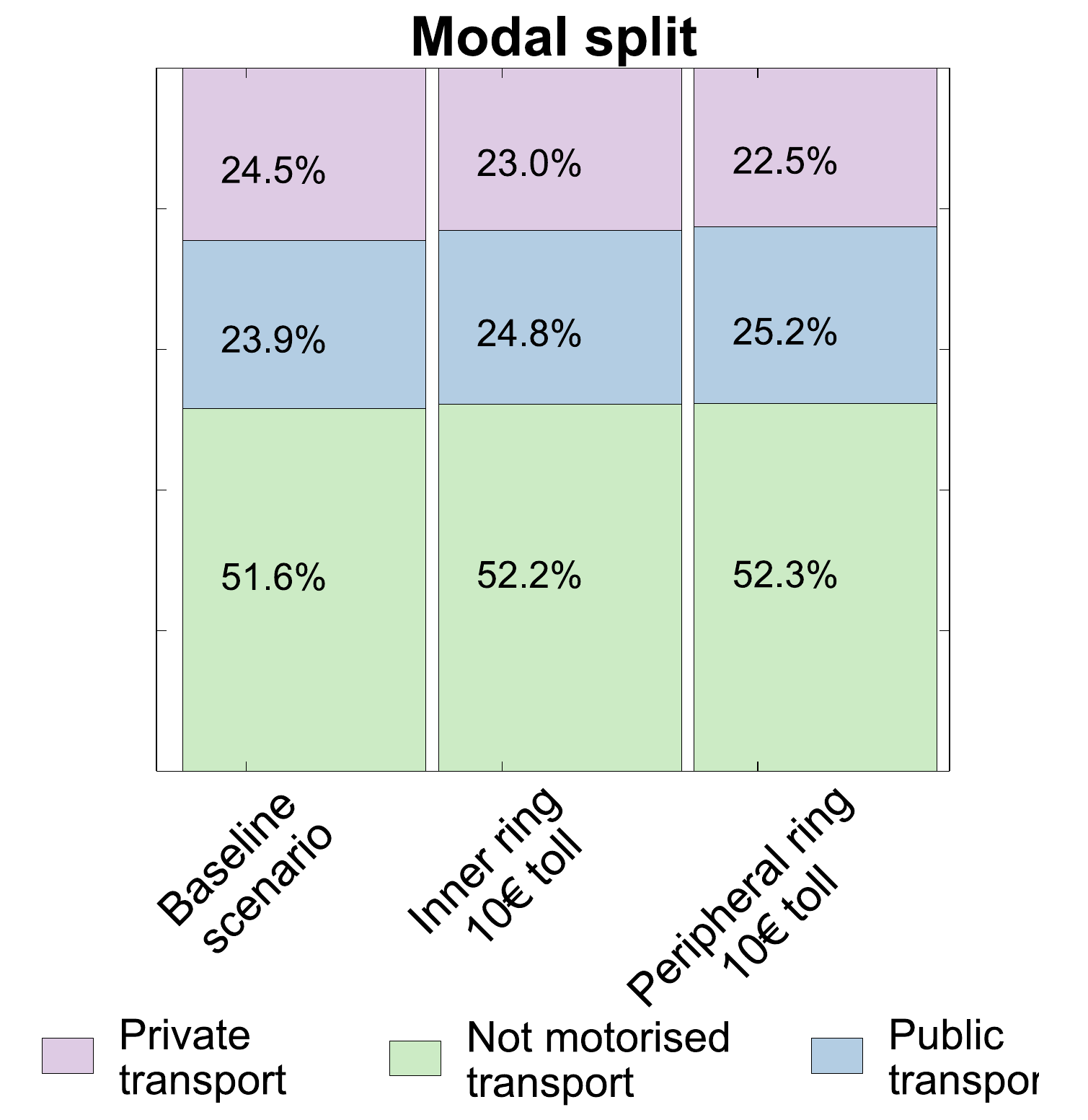}
		\caption{ Comparative of modal split for the region of study  before and after the policy implementation for both the inner and the peripheral ring.\label{fig:Region_modal_split} }
	\end{center}
\end{figure}

\begin{figure}
	\begin{center}
		\includegraphics[width=8cm]{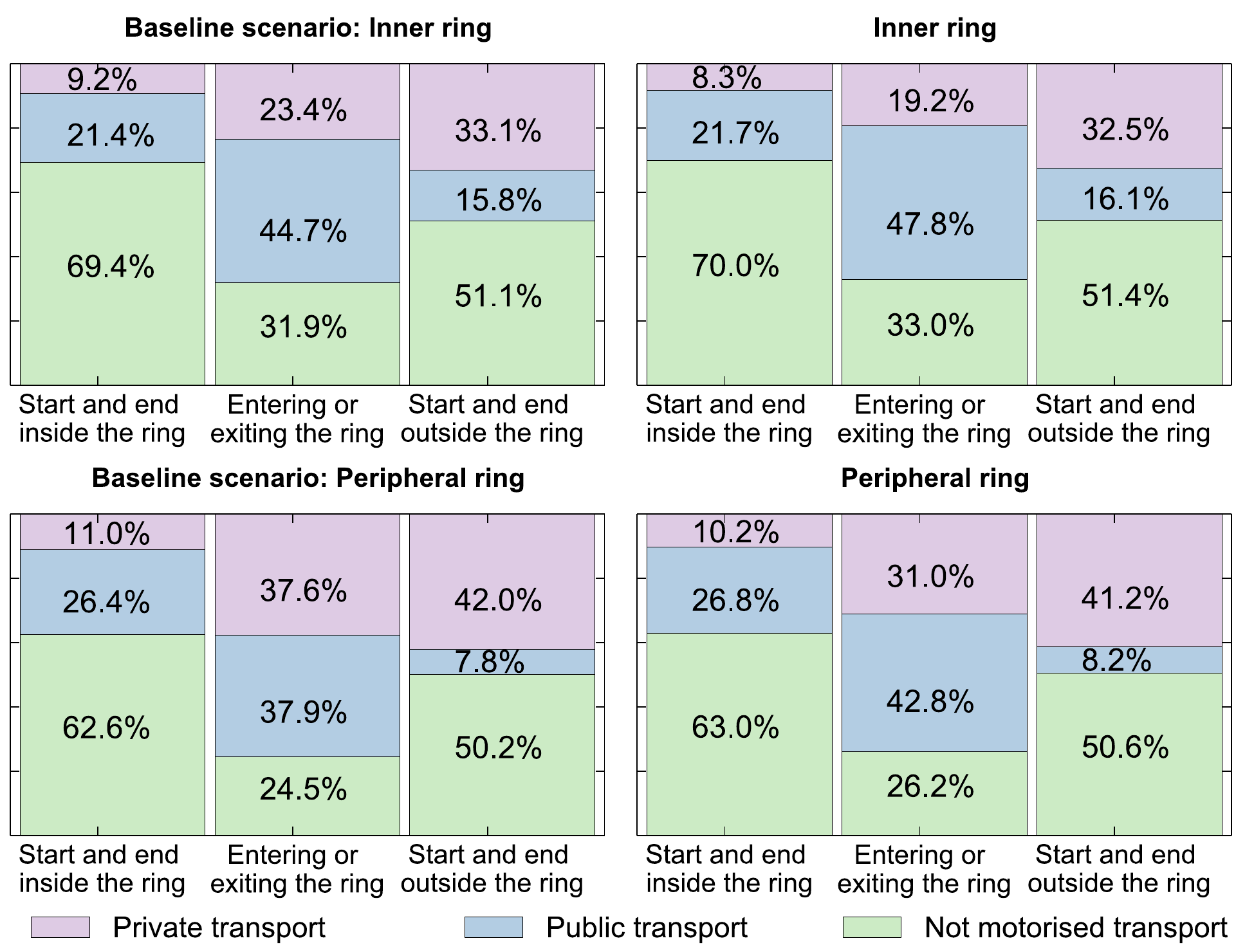}
		\caption{ Modal split for trips starting and ending at different zones for:i a) Baseline scenario considering the different zones separated by the inner ring; b) Policy applied to the inner toll zone, c) Baseline scenario where the different zones are splited by the peripheral ring and d) Policy applied for the peripheral toll zone. For an all day $10$\euro{} charge.
			\label{fig:Modal_split_per_zones} }
	\end{center}
\end{figure}

As the number of total trips by the simulated agents is conserved, the reduction of car trips implies an increase in the use of other modes of transport. Figures \ref{fig:Region_modal_split} and \ref{fig:Modal_split_per_zones} show the resulting modal split before and after the $10$ \euro{} charge application. As can be seen on Figure \ref{fig:Region_modal_split}, most of the car trips have been transferred to public transport. However, this transfer depends on the area where the trips start and end (Figure \ref{fig:Modal_split_per_zones}). For trips starting and ending inside the ring, car trips are almost equally transferred to soft modes (walk and bike) and public transport with a slight preference for soft modes. Car trips crossing the ring in any direction are transferred mainly to public transport. Car trips starting and ending outside the ring are evenly transferred to public transport and soft modes with a slight preference for soft modes in the case of the inner ring and for public transport in the case of the peripheral ring.

Despite the decrease of car traffic, the resulting total average travel time per
trip and average travel time for car trips are not considerably modified.The average travel time for car trips is reduced by 25 seconds, which corresponds to less than $3\%$ of the car trip travel time for the baseline scenario. For the inner ring the higher car trip travel time reduction is observed for trip lengths between $16-25$ $km$ showing a time reduction of around $6\%$  ($1-2$ minutes) and between $70-73$ $km$ with a reduction of around $7-8\%$ ($4.5-6.5$ minutes). In contrast to the inner ring, for the peripheral ring a car trip travel time is increased by around $6\%$ for trips of lengths between $19-26$ $km$. This may be due to trips surrounding the city by the Rondas, which are more congested after the toll implementation. Car travellers that used to cross the city are forced to surround it to avoid the toll charge imposing more congestion to the roads surrounding the toll area. In addition, in the inner ring trips of length between $70-73$ $km$ show the higher travel time reduction with a reduction of around $6-7\%$. The average travel time per trip for all lengths has slightly increased, a bit more than one minute, which corresponds to $5\%$ of the baseline scenario's average travel time. The higher travel time increment is observed for trips of lengths between $12-36$ $km$ showing an increment of $4-8$ minutes (corresponding to $10-20\%$ of the baseline scenario travel time). This may be due to the modal change, since alternative modes are less time efficient than cars.

\subsubsection{Aggregated results for different timing schemes}\label{aggregated results}  

It should be recalled that for the different timing schemes only the $10$\euro{} charge was tested. In terms of net number of car trips the all day charge scheme reports the highest reduction compared with those applied in the morning and afternoon peaks or only in one of them. Total car trips reduction for the morning and afternoon toll charge amounts to $2.24 \%$, very close to that obtained for the all day $5$\euro{} charge (showed in section \ref{all_day}). Similarly, afternoon (morning) toll charge results in a car trip reduction of $1\%$ ($0.8\%$) which is very similar to that of the all day $2$\euro{} charge. This means that the same total number of car trips reduction can be obtained with different combinations of charging times and prices. In terms of revenues raised by the policy application the all day policies generate a higher revenue. The all day $5$\euro{} charge scheme report a $33\%$ more revenues raised than the morning and afternoon charge scheme. And the all day $2$\euro{} charge scheme raise $28\%$ ($20\%$)  more revenues than the morning(afternoon) $10$\euro{} charge scheme.

\begin{figure}
	\begin{center}
		\includegraphics[width=8cm]{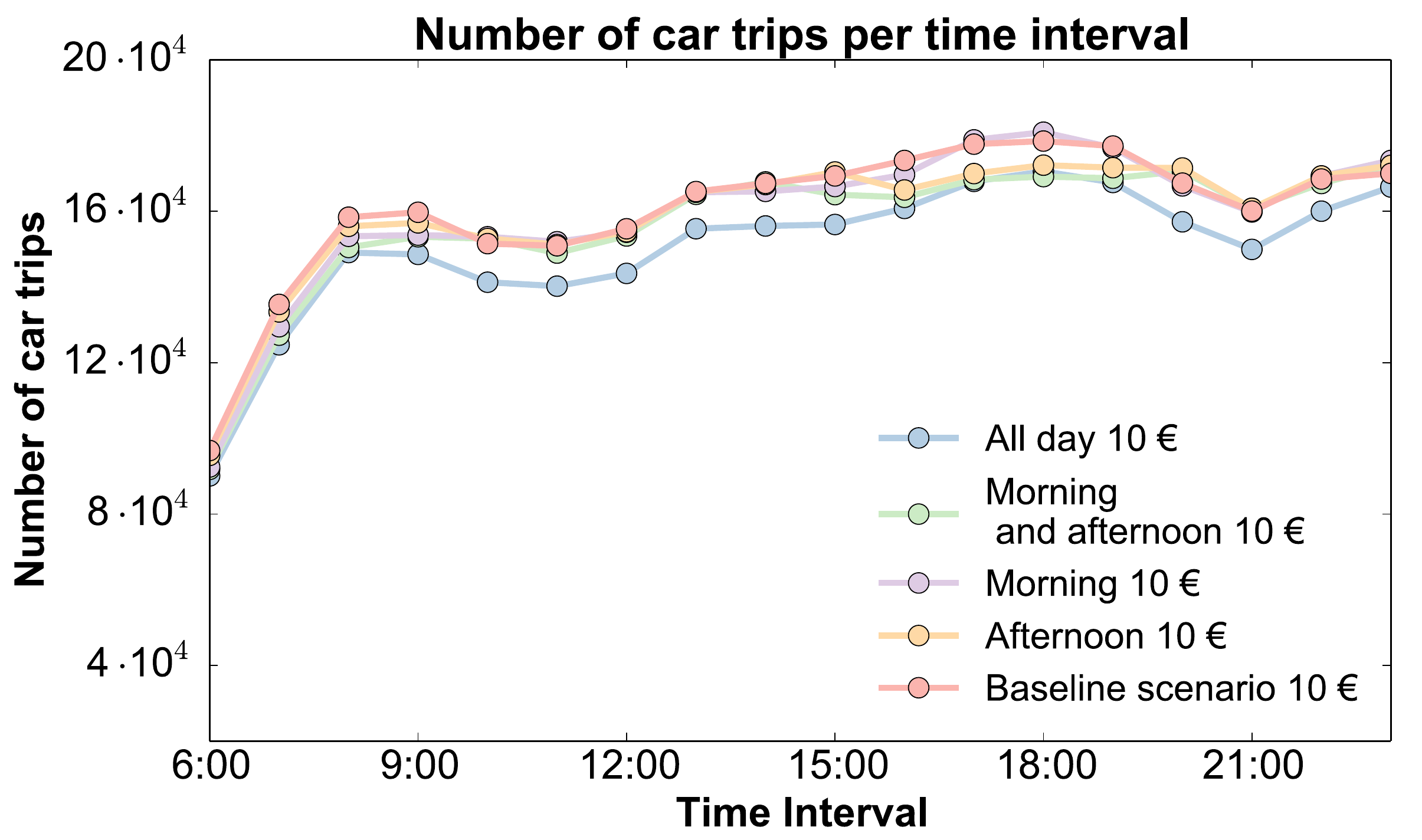}
		\caption{Detail from 6:00 to 22:00 hrs of the number of car trips per time interval for different timing toll schemes.
			\label{trips_time} }
	\end{center}
\end{figure}

Beyond the number of car trips and the revenues raised, the toll policies applied at specific times have some specificities. As a general statement, the effect of toll charge applied at specific times of the day goes beyond the time at which it is applied. In Figure \ref{trips_time}, one can observe that car reduction occurs even for those times where no charge is applied. The toll charge applied during the morning and afternoon peak (form 08:00-10:00 and 16:00-20:00) reduces the car traffic at rush hours not only by reducing the net number of car trips but also by shifting the time at which some of them occur. The dark blue curve on Figure \ref{trips_time} shows that the number of car trips at 8:00 hrs is considerably reduced at the price of being slightly increased at a later time (10:00 hrs) compared with the baseline scenario (red curve). A considerable reduction of trips is also observed between 14:00 and 15:00, before the afternoon charge is applied. This may correspond to return trips of those that decided to leave the car at home to avoid the morning charge or to travellers that decide to leave the car at home in the morning to avoid paying the afternoon charge in their return trip. Some car trips have been shifted from the afternoon peak to just after the toll charges ends, 20:00 hrs. Similar effects are observed for only morning (light blue curve on Figure \ref{trips_time}) and only afternoon (purple curve) toll charges. For the first one a net reduction of car trips with respect to the baseline scenario is observed not only during the charging time but also in the early afternoon, 14:00-16:00, supporting the hypothesis that these trips corresponds to return trips of users who left the car at home to avoid the morning toll. Also some of the trips occurring at the morning peak are shifted to slightly after the toll timing, 11:00 hrs. For the afternoon toll, as expected car trips at the afternoon peak are reduced with respect to the baseline scenario. Most of these trips were shifted to other modes, while a smaller amount were still performed by car but at a later time, once the toll disappears (see the small car trip increment with respect to the baseline scenario occurring just after 20:00 hrs for the purple curve on Figure \ref{trips_time} )

\begin{figure}[b]
	\begin{center}
		\includegraphics[width=8cm]{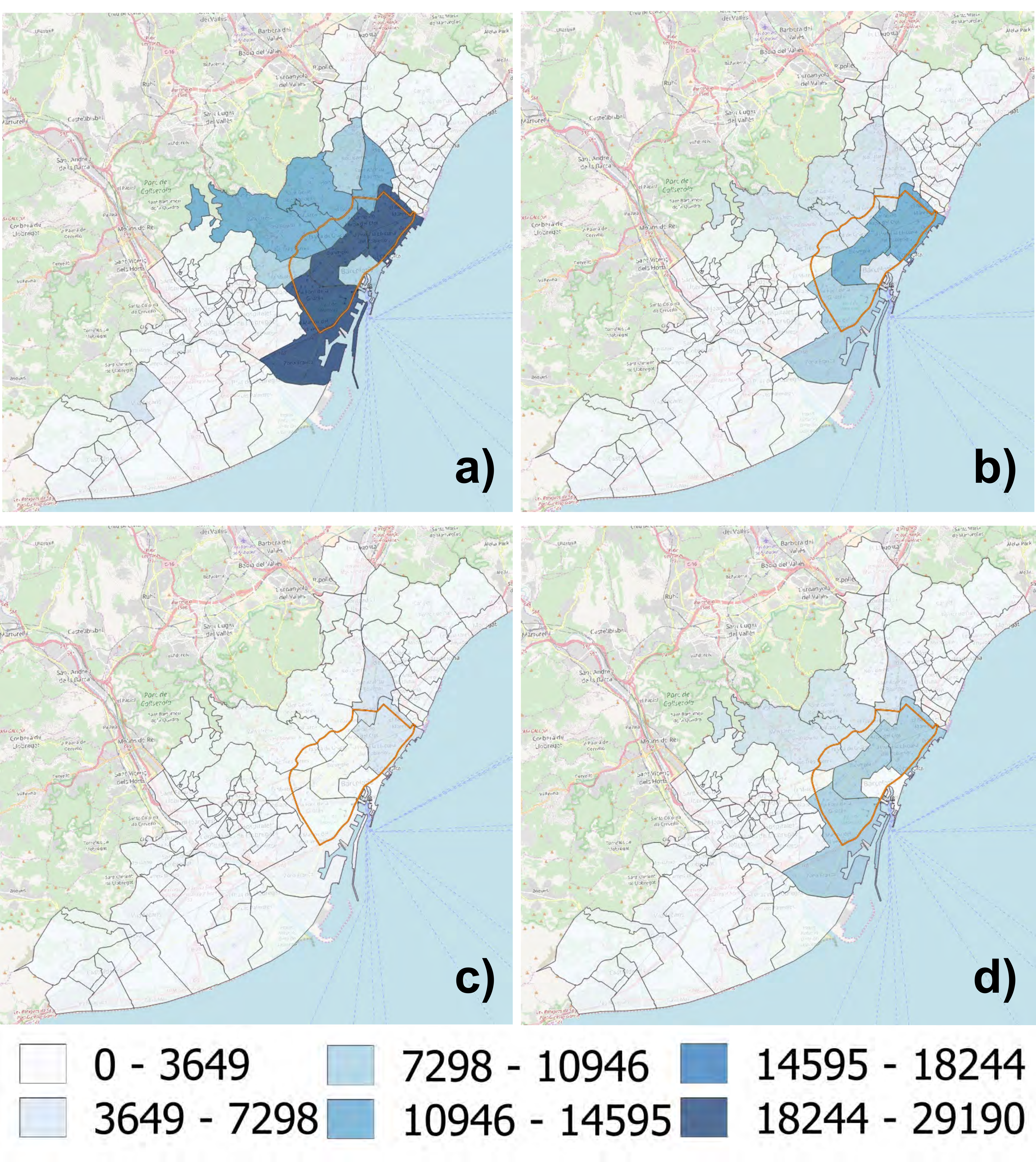}
		\caption{ Number of residents entering by car the area surrounded by the inner ring, before any policy is applied, at: a) any time of the day; b) only during the morning and the afternoon peaks; c) during the morning peak and d) during the afternoon peak. The maps are generated using the standard layout of Open Street Maps. \label{fig:Affected_districts} }
	\end{center}
\end{figure}

\begin{figure}
	\begin{center}
		\includegraphics[width=8cm]{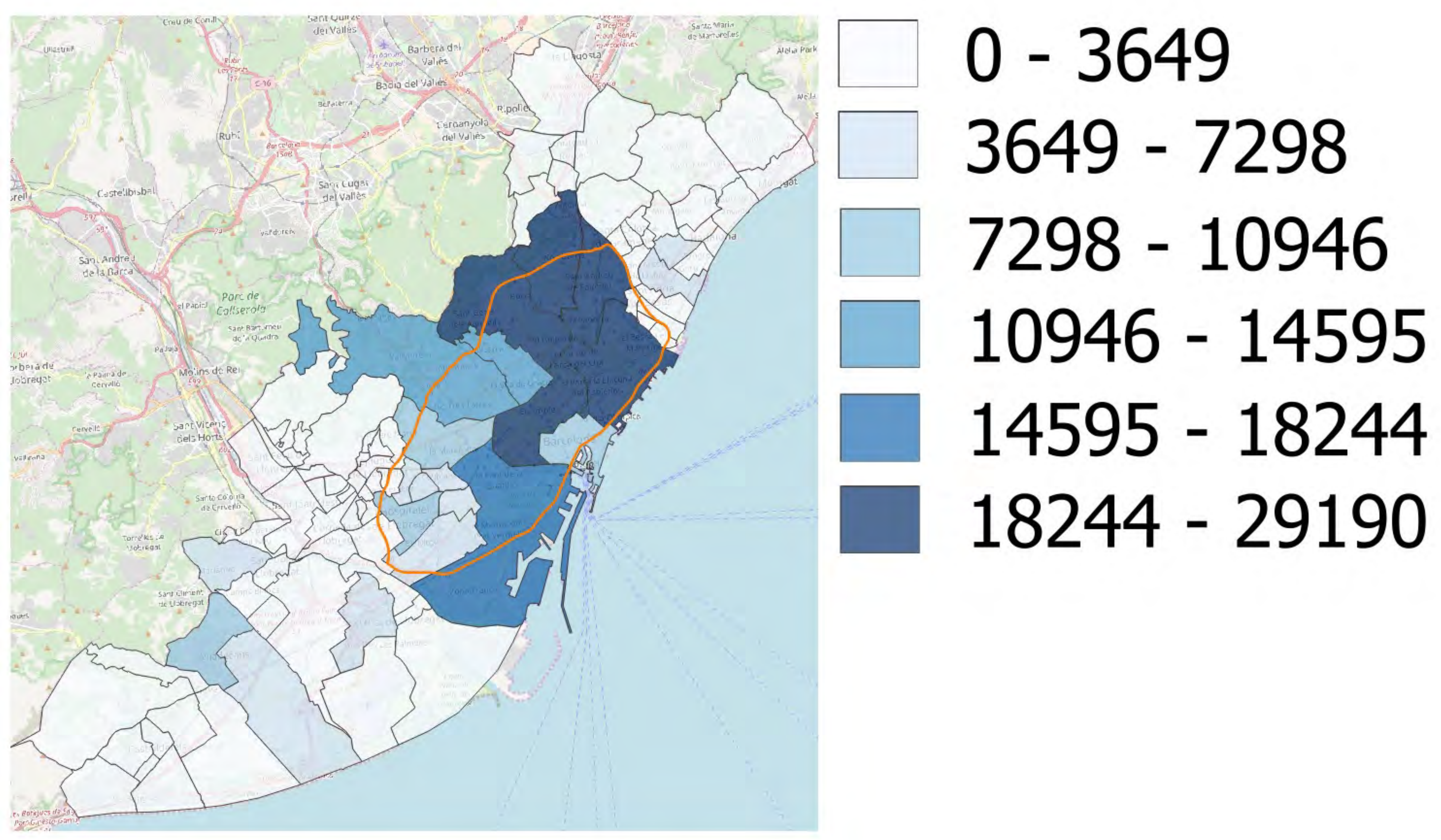}
		\caption{Number of residents entering by car the area surrounded by the peripheral ring, before any policy is applied. The maps are generated using the standard layout of Open Street Maps.\label{fig:Affected_districts_peripheral}}
	\end{center}
\end{figure}

\begin{figure}[b]
	\begin{center}
		\includegraphics[width=8cm]{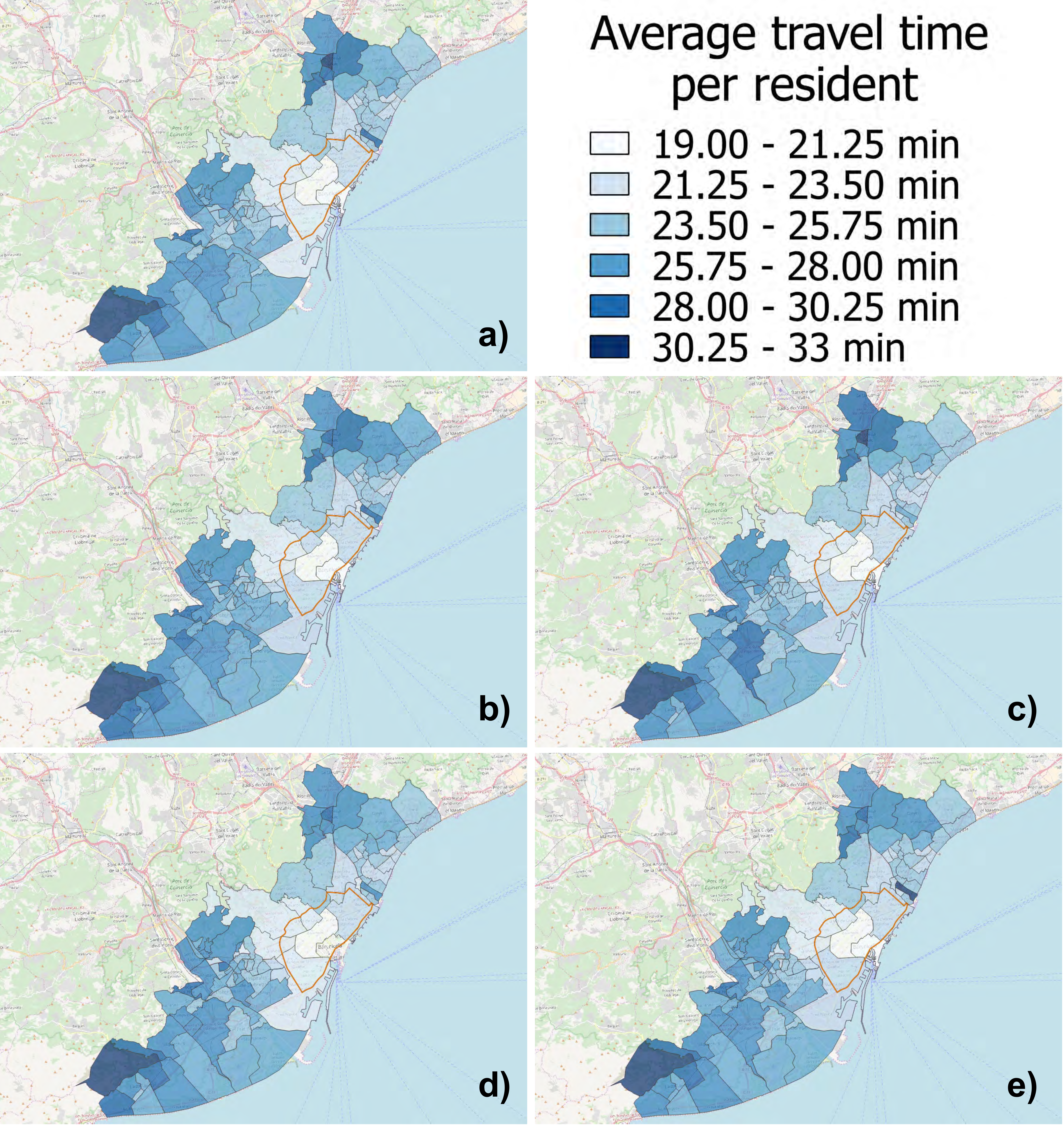}
		\caption{ Average travelled time per district a) Baseline scenario, b) $10$ \euro{} all day toll, c) $10$ \euro{} morning and afternoon toll, d) $10$ \euro{} morning toll and e) $10$ \euro{} afternoon toll. The maps are generated using the standard layout of Open Street Maps.\label{fig:Travel_times} }
	\end{center}
\end{figure}

\begin{figure}
	\begin{center}
		\includegraphics[width=8cm]{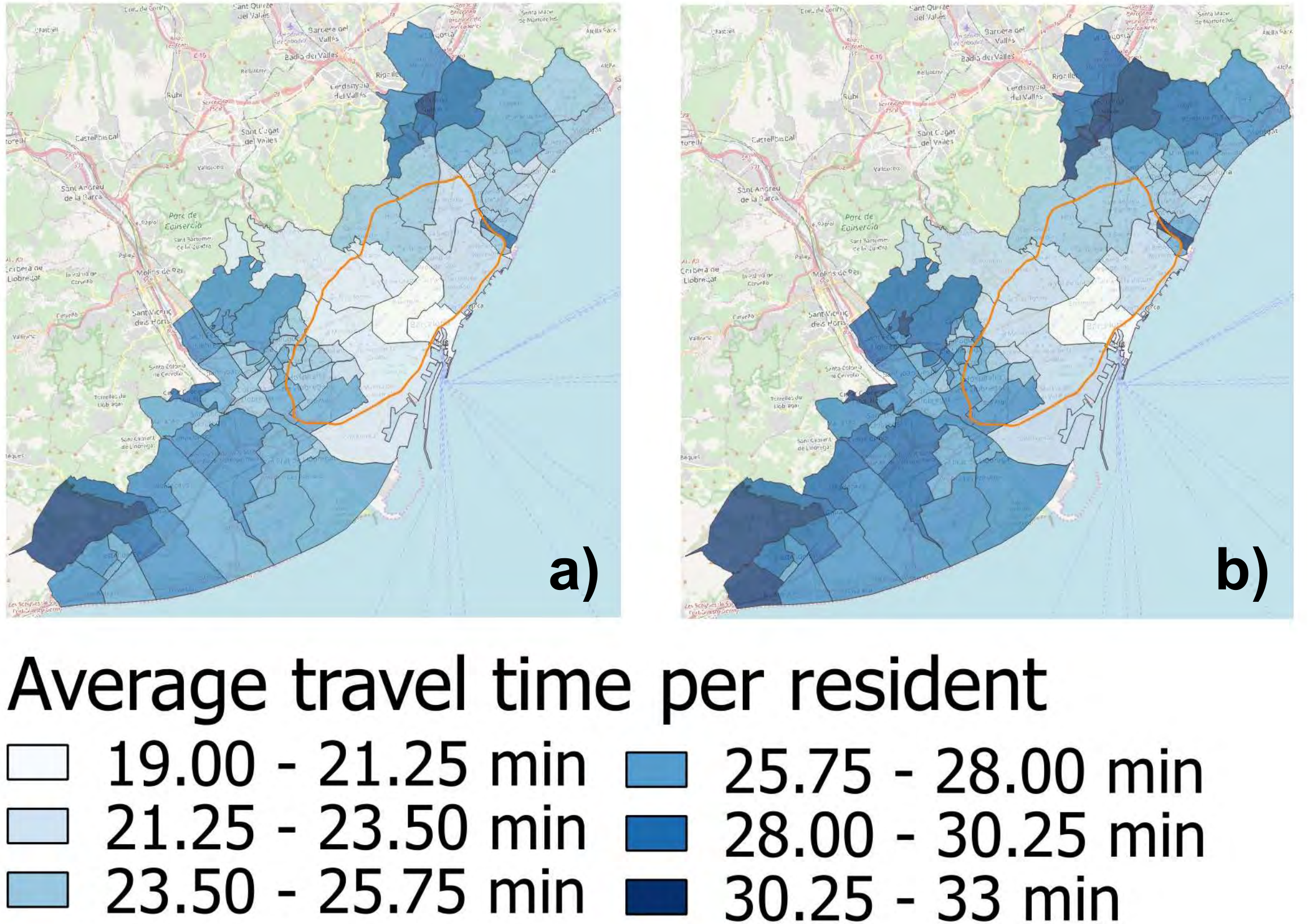}
		\caption{Average travelled time per district. a) Baseline scenario, b) $10$ \euro{} all day toll. The maps are generated using the standard layout of Open Street Maps.\label{fig:Travel_times_ring_2} }
	\end{center}
\end{figure}

\subsubsection{Results from a resident-centric perspective}   

The way a toll charge affects residents of different areas is also influenced by the time at which the toll is applied and by the area covered by the ring. Figures \ref{fig:Affected_districts} and \ref{fig:Affected_districts_peripheral} show the districts that will result more affected by the policy application, in terms of number of residents that usually, before any charge is applied, cross the ring toll by car. It can be seen in \ref{fig:Affected_districts}a and \ref{fig:Affected_districts_peripheral}  that the application of an all day toll policy  virtually affects all districts intersecting the toll area. This is also the case, although to a lesser extent, for the application of the toll at the morning and afternoon peaks (Figure \ref{fig:Affected_districts}b) or at the afternoon peak (Figure \ref{fig:Affected_districts}d) mainly affecting most of the districts intersecting the ring. In a qualitatively different way, the case of applying the toll only during the morning peak (Figure  \ref{fig:Affected_districts}c) would mostly affect the residents of a couple of districts intersecting the toll area.

As discussed in section \ref{aggregated results}, the average travel time is not reduced by the application of the policy. On the contrary, it increases slightly. However, this increment does not affect in the same way residents of the different areas. In terms of travel times, the districts intersecting the inner ring toll seem to be less affected by the policy. The average travel time per district, calculated as the average travel time for those trips performed by the district residents from any origin to any destination, remains unchanged before and after
the policy application regardless the timing scheme (Figure \ref{fig:Travel_times}) and the coverage area of the toll zone (Figure \ref{fig:Travel_times_ring_2}) for these districts. An explanation may be that, although most of the residents of these zones were "forced" to change the transport mode, these districts are well connected by public transport infrastructure. Still, the coverage area of the toll zone and, to a lesser extent, the timing scheme determine which districts are more affected. For the inner ring: all day, morning and afternoon and only morning tolls seem to affect more districts than the only afternoon toll charge (from the comparison of Figure \ref{fig:Travel_times} a with b, c and d). The application of the policy at a larger area, the peripheral ring, affects a greater number of districts including those intersecting the ring toll (comparison of Figure \ref{fig:Travel_times_ring_2}a and b). This may be due to the lower public transport services for areas distant from the city centre.

\begin{figure}
	\begin{center}
		\includegraphics[width=6cm]{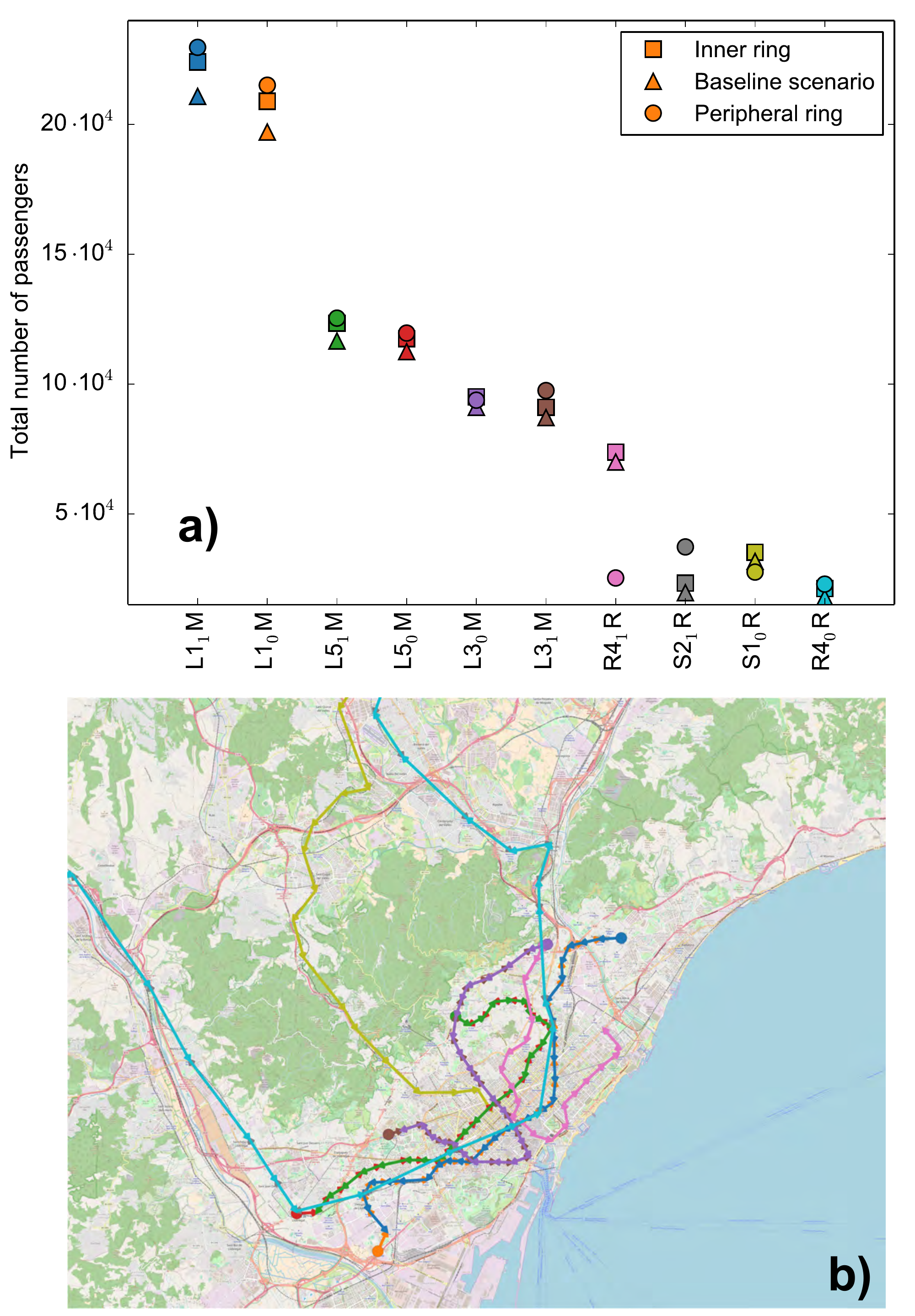}
		\caption{In (a), total number of passengers per line for the train and metro in the baseline scenario ($\triangle$), the inner ring ($\square$) and the peripheral ring $10$ \euro{} scenarios ($\circ$). The lines selected are the ten with the largest change in absolute numbers due to the implementation of the toll policy. Their codes are composed of the number, the direction in subscript ($0$ or $1$) and a letter M for metro or R for train. In (b), map of the metropolitan area displaying the lines. The arrows point to the line direction, while the large circle is the first station. The colour code is maintained in both panels. The maps are generated using the standard layout of Open Street Maps.  }\label{metro_changes}
	\end{center}
\end{figure}

\subsubsection{Impact on the public transport network}\mbox{}\\  
The reduction of car trips is associated to a growth in the demand for public transport. Metro lines are the ones suffering the largest passenger increase after the policy application. For example, in absolute numbers the line most impacted is line $L1$ from Hospital de Bellvitge (West of the city) to Fondo (East) in both directions, which sees its passengers increased in more than $11000$ ($>6 \%$) in each direction (Figure \ref{metro_changes}). In relative numbers, there are lines like the rail R4 going from Sant Vicen\c c de Calders to Manresa that in both directions would see its demand enhanced by $19\%$ in the case of the inner ring toll and over $29\%$ with the peripheral one. Despite these strong increases, the train and metro lines are not expected to get congested thanks to their high frequency and the high capacity of every vehicle.   

To identify which are the most impacted lines and hence need to be reinforced, we focus next on the ratio of load (demand) over capacity. The lines are divided in segments between subsequent stops, and for each segment $i$ we define the normalised load as  
\begin{equation}
\ell_i = \frac{\mbox{Passengers per day in $i$}}{\mbox{Vehicles per day} \times \mbox{Capacity of each vehicle}} .
\end{equation}
The load is then averaged over all the segments of the line to obtain $\langle \ell \rangle$. This metric singles out the lines running over capacity (although since we are taking daily totals it lacks fine temporal resolution). As a transport manager, the most critical lines are those with $\langle \ell \rangle$ largest, especially those close to one. We calculate next the difference between the normalised load of lines in the baseline and the tolled scenarios.

\begin{figure}
	\begin{center}
		\includegraphics[width=6cm]{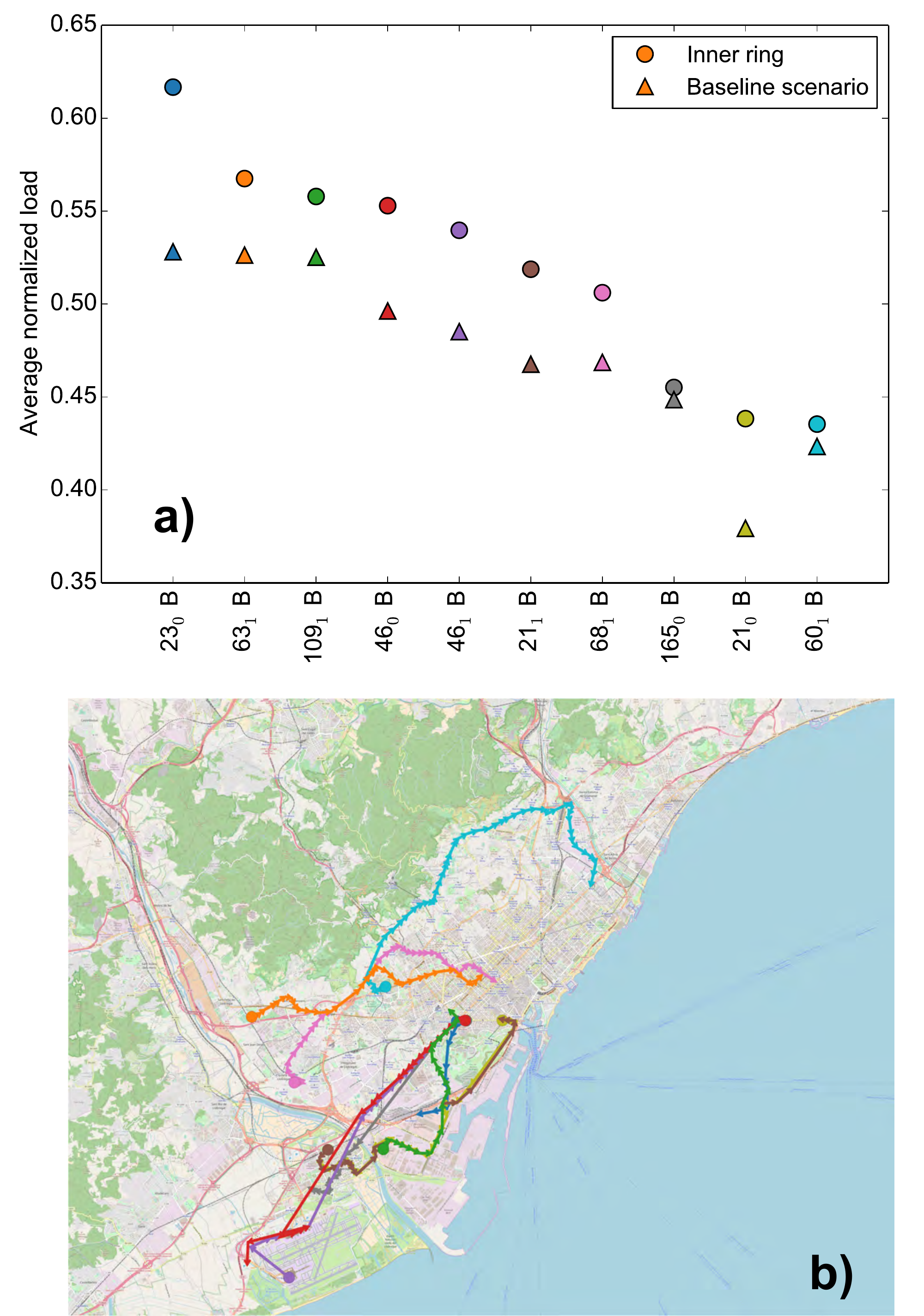}
		\caption{In (a), average normalised load per line in the baseline scenario ($\triangle$) and the inner ring $10$ \euro{} scenario ($\circ$). The lines selected are the ten with the largest change after the implementation of the toll policy. Their codes are composed of the number, the direction in subscript ($0$ or $1$) and a letter B for bus. In (b), map of the metropolitan area displaying the lines. The arrows point to the line direction, while the circle corresponds the first station. The colour code is maintained in both panels. The maps are generated using the standard layout of Open Street Maps. }\label{ring1_pt}
	\end{center}
\end{figure}

\begin{figure}
	\begin{center}
		\includegraphics[width=6cm]{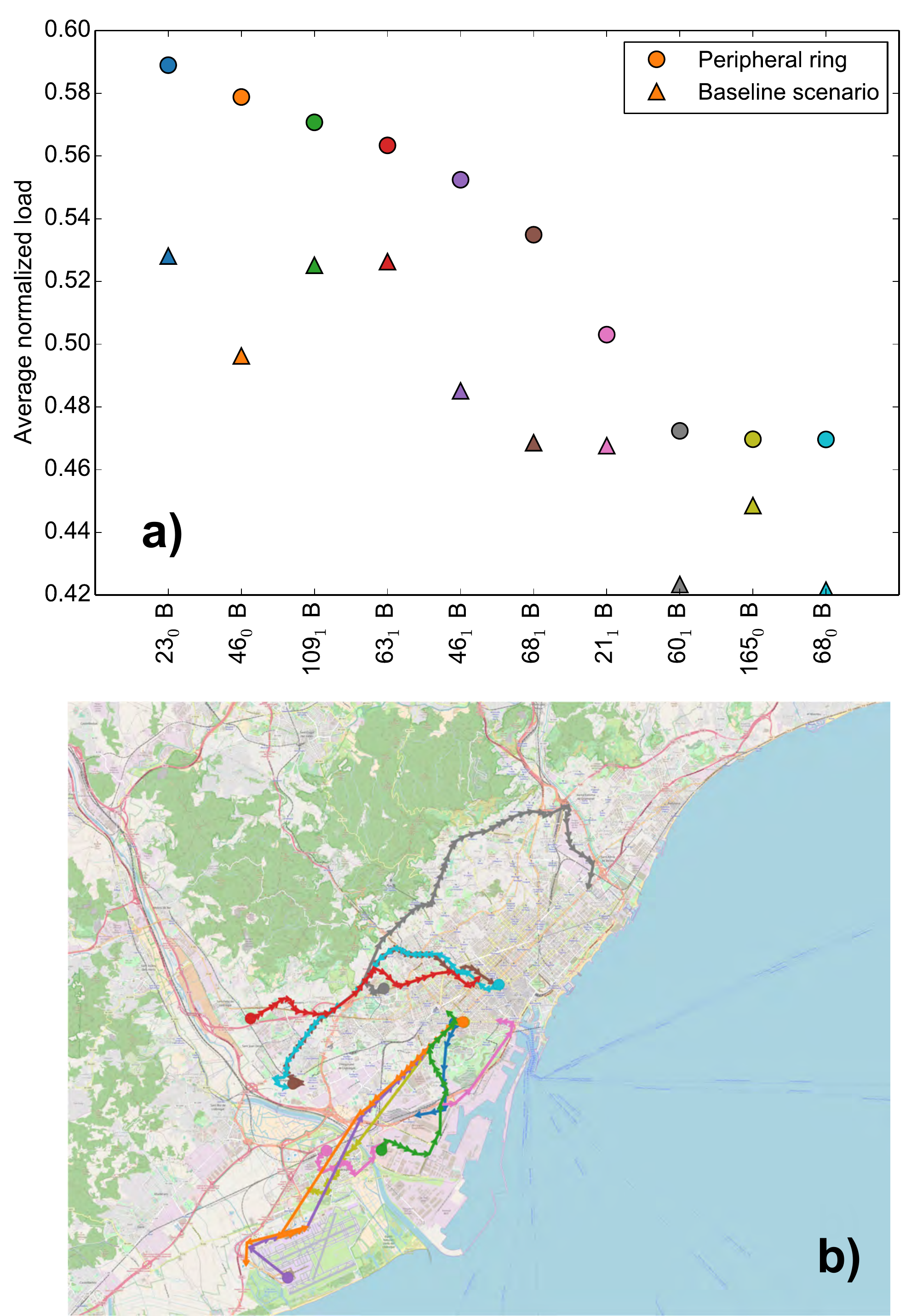}
		\caption{In (a), average normalised load per line in the baseline scenario ($\triangle$) and the peripheral ring $10$ \euro{} scenario ($\circ$). The lines selected are the ten with the largest change after the implementation of the toll policy. Their codes are composed of the number, the direction in subscript ($0$ or $1$) and a letter B for bus. In (b), map of the metropolitan area displaying the lines. The arrows point to the line direction, while the circle corresponds the first station. The colour code is maintained in both panels. The maps are generated using the standard layout of Open Street Maps.}\label{ring2_pt}
	\end{center}
\end{figure}

We studied the effects of the toll policies in detail for all the scenarios but we will focus next on the most interesting results: those for the $10$ \euro{} all day toll. Figures \ref{ring1_pt} and \ref{ring2_pt} show the ten lines with highest $\langle \ell \rangle$ in the network after the toll policy has been applied for both rings. We can see that the ten lines correspond to bus lines (there is no rail or metro). The next fact to highlight is the similarity between the results for both toll scenarios. All the lines connect the centre with the West side of the city. This stresses the high dependency of the West side of Barcelona on the public transport system. Finally, as common result, we find that the lines that suffer the highest increase in its load are those connecting to the airport. 

In the case of the inner ring (Figure \ref{ring1_pt}), we observe that only three lines out of ten do not suffer a significant change after the policy application. One of them (line $60$ in Figure \ref{ring1_pt}) does not cross the ring, while the other two have a high overlap with other more affected lines. In the case of the peripheral ring (Figure \ref{ring2_pt}), we find in the top ranking nearly the same lines as for the inner ring. However, the increase of load compared to the inner ring scheme is significant. One of the most interesting cases is the bus line $60$ in direction North-East  (code $60\_1$), which, unlike in the inner scheme, crosses the ring in the peripheral scheme. This difference causes an increase of $25\%$ of the load in comparison with the scenario with the inner ring. Also note that the line $165$ in direction South, which connects with the West side of the city, emerges in this case in the top ten ranking. Regarding the lines that cross both toll schemes, a significant increase in comparison to the inner ring means that the destinations or origins are located in the region between both toll areas.


\section{Conclusions}\label{sec:conclusions}  

\subsection{Model implementation}

The results obtained for the calibration and validation process support the idea that alternative data sources can be used to meet the data needs of agent-based models and confirm the quality of the activity-travel diaries obtained from mobile phone data. In this context, it is important to stress the good match obtained for the road counts even for a very simple and naif calibration based only on aggregated modal split values. Agent-based models, in this case MATSim, allow us to look at the effects of the policy implementation at an aggregated level but also from a disaggregated, passenger-centric perspective. This includes the observation of side and memory effects not contemplated by aggregated models. For example, a trip going from A to B in MATSim is not seen as a single trip but as part of a full day plan. A person going from B to C cannot do it by car if it has not arrived at B by car from his/her original position A. Similarly, the agent cannot return home by public transport if he/she has arrived at the previous position by car, i.e., all the decisions of travel mode are taken in a single full day plan. This feature may have some drawbacks in exceptional situations, but it reflects the practice in daily mobility.  A policy affecting trips from A to B may have different effects for people performing the same A-B trip depending on their residence place or on the moment of the day at which they need to travel. Therefore, policies applied to the route A-B may have an impact in very different areas of the city, not necessarily only the local ones. The activity-based agent-based nature of MATSim allows such effects to be captured and quantified.

\subsection{Cordon toll policy}

To illustrate these abstract ideas, we have studied the implementation of a toll ring for cars entering in the city of Barcelona. The implementation of the cordon toll policy has positive effects in terms of car use reduction, which may directly affect the level of contaminants and greenhouse gas emissions due to private car usage. The main insights gained from the simulation are the following:
\begin{enumerate}
	\item The intensity of cars trips reduction depends on the toll price and on the area enclosed by the ring toll.
	\item All-day charge schemes seem to be more effective (in terms of car reduction) than charges applied to only specific hours of the day. However, playing with the time of the toll may help to tune the effect of the policy on residents of different areas and shift the traffic peaks.
	\item In terms of number of residents entering the toll zone by car before the policy application, the most affected districts are those intersecting the toll area. In terms of travel time, the most affected districsts are those located farther from the city
	centre, especially for the case of a toll applied to the peripheral ring, which indicate a bad alternative transport option for peripheral areas.
	\item The effects of the cordon toll policy go beyond the tolled zone and the charge period, reducing car trips not only crossing the toll cordon but also modifying those within and outside the toll zone and at travel times different from those of toll application. 
\end{enumerate}

The cordon toll policy does not seem to improve travel times by car in a significant manner, while the average travel time actually increases due to modal shift. This effect could be compensated by earmarking the revenues generated by the congestion pricing scheme for the improvement of public transport. In this sense, some preliminary analysis has been made to see the public transport lines that have received the highest impact (in terms of load raise) as a consequence of the cordon toll policy application.


\section*{Acknowledgement}
This work was developed within the framework of the INSIGHT project funded by the EC under the grant agreement No. 611307. AB is funded by the Conselleria d'Educaci\'o, Cultura i Universitats of the Government of the Balearic Islands and the European Social Fund. AB and JJR also acknowledge partial funding from the Spanish Ministry of Economy, Industry and Competitiveness (MINEICO) and FEDER (EU) under the grant ESOTECOS (FIS2015-63628-C2-2-R).

\newpage

\onecolumngrid
\appendix

\section{Tables}

Aggregated results for both inner and peripheral rings for all toll charges and time schemes tested are presented in the Tables below.

\setcounter{table}{0}

\renewcommand{\tablename}{Table}
\renewcommand{\thetable}{A\arabic{table}}

\begingroup
\squeezetable

\begin{table}[h!]
\begin{center}
\begin{tabular}{| l | l | l | l | l | l | l | }
		\hline	
		Peripheral Ring & Baseline scenario & Toll 2 \euro{}  & Toll 5 \euro{}  & Toll 10 \euro{} \\ \hline
		Total number of trips & 10,368,560 & 10,350,070 & 10,368,850 & 10,366,690\\ \hline
		Total number of car trips & 2,541,700 & 2,497,220 & 2,427,580 & 2,336,560\\ \hline
		Total travel distance (km) & 55,471,436& 55,568,329& 55,578,228& 55,784,940\\ \hline
		Average travel distance& & & & \\
		per trip (km) & 5.35 & 5.37 & 5.38 & 5.38\\ \hline
		Total travel distance  by car  (km) & 28,954,043& 28,118,672 & 28,026,161 & 27,101,526\\ \hline
		Average travel distance  by car  & 11,392 & 11,260 & 11,545 & 11,599\\ \hline
		Average travel time & 00:23:09 & 00:24:17 & 00:23:45 & 00:24:17\\ \hline
		Average travel time by car & 00:14:15 & 00:13:26 & 00:14:25   & 00:14:41\\ \hline
		Number of users paying toll & 0  & 378,200 & 350,490   & 309,470 \\ \hline
		Number of trips paying toll & 0 & 504,040 &  442,750  & 382,690 \\ \hline
		Car trips in the toll zone & 604,350 &  671,390 & 562,120 & 582,370\\ \hline
		Average travel time for trips & & & & \\
		in the toll zone & 00:18:57  & 00:20:07 & 00:19:05  & 00:19:10\\ \hline
		Average travel time for car & & & & \\
		trips in the toll zone & 00:04:49 & 00:05:05 & 00:04:51 &  00:04:52\\ \hline
		Car trips entering the toll zone & 396,040 & 366,270 & 356,220 & 325,350\\ \hline
		Average travel time for trips & & & & \\
		entering the toll zone & 00:33:45 & 00:35:34 & 00:36:44 & 00:38:53\\ \hline
		Average travel time for car & & & & \\
		trips entering the toll zone & 00:19:54 & 00:19:42 & 00:23:06 & 00:25:14\\ \hline
		Car trips exiting the toll zone & 396,490 & 367,390 & 357,710 & 327,400\\ \hline
		Average travel time for trips & & & & \\
		exiting the toll zone & 00:34:13 & 00:36:22 &  00:36:08 & 00:38:03\\ \hline
		Average travel time for car & & & & \\
		trips exiting the toll zone & 00:19:59 & 00:19:05 & 00:19:37 & 00:20:19\\ \hline
		Car trips occurring outside the & & & & \\
		toll zone & 1,144,820 & 1,092,170 & 1,131,280 & 1,121,690\\ \hline
		Average travel time for trips & & & & \\
		occurring outside the toll zone & 00:24:41 & 00:24:52 & 00:36:22 & 00:25:00\\ \hline
		Average travel time for car trips & & & & \\
		occurring outside the toll zone & 00:15:18 & 00:14:33 & 00:14:58 & 00:14:55\\ \hline		
	\end{tabular}
	\caption{Aggregated results for the peripheral ring}
		\label{table_peri}
\end{center}
\end{table}

\endgroup

\newpage

\begingroup
\squeezetable


\begin{table}
\begin{center}

  \begin{tabular}{ | l | l | l | l | l | l | l | l |}
    \hline
     Inner Ring & Baseline scenario & Toll 2 \euro{}  & Toll 5 \euro{}  & Toll 10 \euro{}  & Toll 10 \euro{} morning & Toll 10 \euro{} & Toll 10 \euro{}  \\
	&  & all day   & all day  & all day & and afternoon & morning &afternoon  \\ \hline
Total number of trips & 10,368,560  &  10,371,190  & 10,368,940  & 10,368,170 & 10,369,640  & 10,372,050  & 10,369,910  \\ \hline
Total number of car trips & 2,541,700 & 2,510,470 & 2,458,000 & 2,384,000 & 2,484,540 & 2,521,250 & 2,515,220\\ \hline
Total travel distance (km) & 55,471,436 & 55,562,930 & 55,585,261 & 55,596,113 & 55,543,099 & 55,548,521& 55,503,081\\ \hline
Average travel distance per trip (km) & 5.35 & 5.36 & 5.36 &  5.36 &  5.36 & 5.36 & 5.35 \\ \hline
Total travel distance by car (km) & 28,954,043 & 28,743,170 & 28,275,243 &  27,711,234 & 28,521,102& 28,799,604 & 28,736,127 \\ \hline
Average travel distance & & & & &  &  &\\
by car mode per trip (km)  & 11,392 & 11,449 & 11,503 & 11,624 & 11,479 & 11,423 & 11,425\\ \hline
Average travel time per trip & 00:23:09 & 00:23:17 & 00:23:28 & 00:23:43 & 00:23:40 & 00:23:34 & 00:23:36\\ \hline
Average travel time& & & & &  &  &\\
per trip by car & 00:14:15 & 00:14:14 & 00:14:04 & 00:14:04 & 00:14:02 & 00:14:10 & 00:14:10\\ \hline
Average travel time per trip  & & & & &  &  &\\
by car in the morning peak  & 00:15:46 & 00:15:41 & 00:15:39 & 00:15:36 & 00:15:42 & 00:15:38 & 00:15:51\\ \hline
Average travel time per trip & & & & &  &  &\\
by car mode in the afternoon peak& 00:16:23 & 00:16:21 & 00:15:58 & 00:16:00 & 00:15:32 & 00:16:19 & 00:15:37\\ \hline
Number of users paying toll & 0 & 319,300 & 292,260 & 264,430 & 112,390 & 57,970 & 63,560\\ \hline
Number of trips paying toll & 0 & 411,810 & 363,410 & 322,270 & 121,820 & 59,110 & 65,440\\ \hline
Car trips occurring within & & & & &  &  &\\
the toll zone & 254,510 & 248,190 & 242,580 & 229,470 & 244,400 & 251,150 & 249,450\\ \hline
Average travel time for & & & & &  &  &\\
within toll zone trips & 00:15:37 & 00:15:40 & 00:15:44 & 00:15:47 & 00:15:42 &00:15:40 & 00:15:41\\ \hline
Average travel time for car & & & & &  &  &\\
trips within the toll zone & 00:03:24 & 00:03:24 & 00:03:23 & 00:03:23 & 00:03:25 & 00:03:23 & 00:03:24\\ \hline
Car trips entering the toll zone & 276,910 & 264,620 & 247,640 & 225,880 & 257,340 & 267,990 & 266,720\\ \hline
Average travel time for trips& & & & &  &  &\\
entering the toll zone & 00:30:11 & 00:30:41 & 00:31:15 & 00:32:11 & 00:30:50 & 00:30:31 & 00:30:29\\ \hline
Average travel time for car& & & & &  &  &\\
trips entering the toll zone & 00:17:18 & 00:17:26 & 00:17:28 & 00:17:45 & 00:16:59 & 00:17:26 & 00:16:52\\ \hline
Car trips exiting in the toll& & & & &  &  &\\
zone & 277,200 & 265,160 & 248,410 & 227,000 & 257,660 & 268,670 & 267,360\\ \hline
Average travel time for trips& & & & &  &  &\\
exiting the toll zone & 00:30:33 & 00:30:57 & 00:31:34 & 00:32:34 & 00:31:19 & 00:30:49 & 00:30:58 \\ \hline
Average travel time for car& & & & &  &  &\\
trips exiting the toll zone & 00:17:33 & 00:17:34 & 00:17:23 & 00:17:33 & 00:17:28 & 00:17:23 & 00:17:40\\ \hline
Car trips occurring outside& & & & &  &  &\\
the toll zone & 1,733,080 & 1,732,500 & 1,719,370  & 1,701,650 & 1,725,140 & 1,733,440 & 1,731,690\\ \hline
Average travel time for trips occurring & & & & &  &  &\\
outside the toll zone  & 00:24:35 & 00:24:36 & 00:24:39 & 00:24:41 & 00:24:33  & 00:24:33 &  00:24:34\\ \hline
  \end{tabular}
  \caption{Aggregated results for the inner ring}
  \label{table_inner}
  \end{center}
  \end{table}
  \endgroup

\end{document}